\documentclass[12pt]{article}
\usepackage{amsmath,amssymb,amsfonts}
\usepackage{epsfig}

\usepackage[vcentermath]{youngtab}

\makeatletter \@addtoreset{equation}{section} \makeatother

\renewcommand{\baselinestretch}{1.2}
\begin{document}

\renewcommand{\thefootnote}{\alph{footnote}}

\begin{titlepage}

\begin{center}
\hfill {\tt SNUTP10-006}\\
\hfill{}

\vspace{3cm}

{\large\bf Semi-classical monopole operators in Chern-Simons-matter
theories}

\vspace{2cm}

\renewcommand{\thefootnote}{\alph{footnote}}

{\large Hee-Cheol Kim and Seok Kim}

\vspace{1cm}

\textit{Department of Physics and Astronomy, Seoul National
University, Seoul 151-747, Korea.\\ Center for Theoretical Physics,
Seoul National University, Seoul 151-747, Korea.}

\vspace{0.7cm}

E-mails: {\tt heecheol1@gmail.com, skim@phya.snu.ac.kr}

\end{center}

\vspace{2.5cm}

\begin{abstract}

We construct classical solutions for magnetic monopole
operators in $\mathcal{N}\!=\!6$ superconformal Chern-Simons-matter
theories. In particular, we explicitly find solutions with unequal magnetic
flux contents in the two $U(N)$ gauge groups, whose existence has been known
only indirectly. We also study the ground state degeneracies of the
$U(2)\times U(2)$ monopoles by quantizing the moduli of the solution.

\end{abstract}

\end{titlepage}

\renewcommand{\thefootnote}{\arabic{footnote}}

\setcounter{footnote}{0}

\renewcommand{\baselinestretch}{1}

\tableofcontents

\renewcommand{\baselinestretch}{1.2}

\section{Introduction}

A class of Chern-Simons-matter theories provide a way of microscopically
studying M2-branes and M-theory (via gauge/gravity duality), as explored extensively
in recent days after \cite{Bagger:2006sk,Gustavsson:2007vu,Aharony:2008ug}.
The simplest model is perhaps the $\mathcal{N}\!=\!6$ superconformal Chern-Simons-matter
theory with $U(N)_k\times U(N)_{-k}$ gauge group and Chern-Simons level,
which describes the dynamics of $N$ M2-branes probing
$\mathbb{R}^8/\mathbb{Z}_k$ \cite{Aharony:2008ug}.

One of the key ingredients which makes the Chern-Simons-matter
theory possible to describe M2-branes and M-theory is the magnetic monopole operators
\cite{Borokhov:2002ib}. Firstly, such operators play crucial roles in providing
the correct spectrum of local operators to account for the states in the dual gravity \cite{Aharony:2008ug,Berenstein:2008dc,Kim:2009wb}. Monopole operators
are also expected to play a central role in the supersymmetry enhancement from
$\mathcal{N}\!=\!6$ to $\mathcal{N}\!=\!8$ when the Chern-Simons level $k$ is
$1$ or $2$, as studied in \cite{Aharony:2008ug,Benna:2009xd,Gustavsson:2009pm,Kwon:2009ar}
from various viewpoints. In particular,
it has been shown in \cite{Kim:2009wb} that the spectrum of protected operators
including monopoles perfectly matches with that of the gravity dual, including the cases
with $k=1,2$. Since the gravity spectrum in the latter cases is tightly organized by
$\mathcal{N}\!=\!8$ supersymmetry, the agreement checked in \cite{Kim:2009wb} provides
a strong support of supersymmetry enhancement including monopoles.


Local operators in conformal field theories are in 1-to-1 correspondence to the
states in the radially quantized theories. Local operators in $\mathbb{R}^3$
containing monopoles therefore map to states in $S^2\times\mathbb{R}$ with
nonzero magnetic flux on $S^2$. From the field theory perspective, monopole operators
are non-perturbative objects in its coupling constant $\frac{1}{k}$. For large enough
$k$, it is therefore natural and technically feasible to study classical `solitonic'
solutions on $S^2\times\mathbb{R}$ with nonzero monopole charges.

In this paper, we study classical monopoles in the radially quantized $\mathcal{N}\!=\!6$
Chern-Simons-matter theory preserving minimal number of supercharges (2 Hermitian).
With large $k$, we can semi-classically quantize these solutions and study quantum aspects
of the monopole operators which have been addressed by more indirect methods
\cite{Kim:2009wb,Benna:2009xd}.

Monopole operators inserted at a point in 3 dimension create a magnetic flux in a $U(1)$
subgroup of the gauge group, on a 2-sphere surrounding the insertion point. In the
$\mathcal{N}\!=\!6$ Chern-Simons-matter theory,
the $U(1)$ subgroup is chosen by specifying two diagonal matrices with integer entries,
\begin{equation}\label{monopole-charge}
  H={\rm diag}(n_1,n_2,\cdots,n_N)\ ,\ \
  \tilde{H}={\rm diag}(\tilde{n}_1,\tilde{n}_2,\cdots,\tilde{n}_N)
\end{equation}
where the entries can be ordered to be non-increasing $n_1\geq\cdots\geq n_N$,
$\tilde{n}_1\geq\cdots\geq\tilde{n}_N$ using the Weyl group. From the structure of this
theory, these fluxes are subject to the constraint $\sum_{i=1}^Nn_i=\sum_{i=1}^N\tilde{n}_i$.
These integers (partly) specify the boundary behaviors of the gauge fields around
the point at which the monopole operator is inserted:
\begin{equation}
  \frac{1}{2\pi}\int_{S^2}F=H\ ,\ \ \frac{1}{2\pi}\int_{S^2}\tilde{F}=\tilde{H}
\end{equation}
where $F,\tilde{F}$ are the $U(N)\times U(N)$ field strengths. As we shall explain
in this paper, the field strengths $F$ and $\tilde{F}$ do not have to be (and in certain
cases, cannot be) uniform on the 2-sphere.

Monopole operators with magnetic charges satisfying
$H=\tilde{H}$ have been studied quite extensively in the literature \cite{Aharony:2008ug,Berenstein:2008dc,Klebanov:2008vq,Benna:2009xd}.
This type of monopole operators is important in that they support gauge invariant
chiral operators, whose scale dimensions are given by their R-charges. A detailed
semi-classical study of such monopole operators for large $k$ has been done in \cite{Berenstein:2009sa,Kim:2009ia}. In particular, the semi-classical
solutions corresponding to the chiral operators are fairly simple. This is because
one only has to excite the s-waves of the scalar matters on $S^2$, and the excited matters
always come in the diagonal form. The excitations around
such semiclassical solutions have been studied in detail in \cite{Berenstein:2009sa,Kim:2009ia},
from which (the index version of) the partition function for these states
\cite{Kim:2009wb} was calculated in a direct way \cite{Kim:2009ia}.

It has been noticed that general type of monopole operators with
$H\neq \tilde{H}$ should also exist. Most importantly, they are required
to have the large $N$ spectrum of the protected operators agree with
that of the gravity states \cite{Kim:2009wb}. Such monopoles have also been studied in
\cite{Imamura:2009hc,Kim:2010vw}. However, a direct understanding of such monopole
operators from the Chern-Simons-matter theory is lacking.

In this paper, we provide an `honest construction' of classical monopole solutions with
$H\neq\tilde{H}$ by constructing their classical solutions. While chiral operators
preserve at least
$4$ real supercharges, the solutions with $H\neq\tilde{H}$ generically preserve $2$
real supercharges, which makes the structure of the classical solutions more delicate
than the former ones. For instance, monopole operators with $H\!\neq\tilde{H}$
always come with nonzero angular momentum \cite{Kim:2009wb}, which makes
the classical solutions to carry nontrivial angular dependence on $S^2$.
In particular, contrary to the monopoles studied in the 3 dimensional SQED
\cite{Borokhov:2002ib} or in Chern-Simons-matter theories for chiral operators
\cite{Berenstein:2009sa,Kim:2009ia}, ground state solutions for Chern-Simons
monopoles with $H\!\neq\!\tilde{H}$ carry non-uniform magnetic flux on $S^2$ due
to the backreaction of matters not in s-waves.

We shall mostly restrict
our studies to the monopoles in the $U(2)\times U(2)$ theory, although some special
solutions for the $U(N)\times U(N)$ group will also be presented.

Another interesting subject is the degeneracy of ground states for
given set of monopole charges $H,\tilde{H}$. For $H=\tilde{H}$, the degeneracy is determined
by the study of chiral rings with monopole operators \cite{Aharony:2008ug,Hanany:2008qc}.
From the viewpoint of classical solutions, this is obtained by quantizing the classical
moduli space of the solution \cite{Kim:2009ia}. The data on the ground state
degeneracy for monopoles with $H\!\neq\tilde{H}$ is encoded in the index of \cite{Kim:2009wb}.
In this paper, we explicitly obtain the ground state
degeneracies for various magnetic charges by quantizing our classical solutions.
For the case with $H=(n_1,0)$, $\tilde{H}=(\tilde{n}_1,\tilde{n}_2)$, namely when one of the
$U(2)\times U(2)$ fluxes is zero, we find that the ground state degeneracy calculated from our
solution completely agrees with the result from the index, implying that our classical solution
is most general.\footnote{Although $H$, $\tilde{H}$ are diagonal matrices,
we shall often write them simply as integer sequences for brevity.}
On the other hand, for general $U(2)\times U(2)$ magnetic charges, the degeneracy
obtained from our solution is smaller than that obtained from the index. This implies that
a more general ansatz than ours is necessary for the most general solution. See
section 2 for the form of our ansatz, and conclusion as well.

The remaining part of this paper is organized as follows. In section 2, we
summarize the aspects of monopole operators found in \cite{Kim:2009wb} that
we would like to understand more directly. In particular, we explain the spectrum
and ground state degeneracies of monopole operators with various magnetic charges.
In section 3, we construct classical solutions for monopole operators with
$H\!\neq\!\tilde{H}$ starting from an ansatz. The monopoles with $U(1)\times U(2)$ gauge
group are considered first in section 3.1 as they exhibit relatively simple behaviors,
which is then extended to the general $U(2)\times U(2)$ fluxes in
section 3.2. In all cases, we find a set of ordinary differential equations,
whose solutions are obtained numerically. We also semi-classically quantize
the moduli of the solutions, obtain the ground state degeneracies and compare with the results
of section 2. Some special solutions for larger gauge groups are also presented in section 3.3.
Section 4 concludes with discussions. Appendix A explains the monopole solutions in a simple
$U(1)$ Chern-Simons-matter theory, to illustrate that the type of solutions we study
in this paper is common in all Chern-Simons-matter theories.

\section{Ground states of monopole operators from the index}

The index for local gauge invariant operators preserving a particular set of
$2$ real supercharges (or one complex and its conjugate)
in the $\mathcal{N}\!=\!6$ superconformal Chern-Simons-matter
theories was computed in \cite{Kim:2009wb}.
As one takes the Cartans of the $SO(6)$ R-symmetry to be three $U(1)$'s which rotate the
three `orthogonal 2-planes,' the chosen supercharge is charged under one of them which we
call $q$, while being neutral under the other two. Let us denote the latter two
charges by $q_1,q_2$, which
are Cartans of $SO(4)\subset SO(6)$. The index takes the form of \cite{Kinney:2005ej}
\begin{equation}
  I(x,y_1,y_2)={\rm Tr}\left[(-1)^Fx^{\epsilon+j}y_1^{q_1}y_2^{q_2}\right]\ ,
\end{equation}
where the trace is taken over the space of local gauge invariant operators,
$F$ is the fermion number of the local operators, $\epsilon$ is the scale dimension
(or the energy of states in the radially quantized theory), $j$ is the Cartan of
the $SO(3)$ angular momentum on $\mathbb{R}^3$. This index counts the local operators
whose dimensions saturate the BPS bound $\epsilon=q+j$.

An integral expression of this index was obtained in \cite{Kim:2009wb}. The
index acquires contribution from sectors with various magnetic
monopole charges $H,\tilde{H}$ of the form (\ref{monopole-charge}).
One of the main interests in this paper is to study local operators with smallest scale
dimensions for given magnetic charges: in other words, we are interested in the
`ground states.' Therefore, let us explain the reduction of the index
which contains information on the ground state spectrum and degeneracy for given
monopole charge. The relevant expression is given by
\begin{equation}\label{index}
  \hspace*{-0.5cm}I_{H,\tilde{H}}(x)\!=\!\frac{x^{\epsilon_0}}
  {(\rm symmetry)}\!\int\!\!\prod_{i=1}^N\!
  \left[\frac{d\alpha_id\tilde\alpha_i}{(2\pi)^2}\right]\!e^{ik\sum_{i=1}^N(n_i\alpha_i
  \!-\!\tilde{n}_i\tilde\alpha_i)}
  \frac{\prod_{i\neq j}(1\!-\!x^{|n_i\!-\!n_j|}e^{-i(\alpha_i\!-\!\alpha_j)})
  (1\!-\!x^{|\tilde{n}_i\!-\!\tilde{n}_j|}e^{-i(\tilde\alpha_i\!-\!\tilde\alpha_j)})}
  {\prod_{i,j}(1\!-\!rx^{\frac{1}{2}\!+\!|n_i\!-\!\tilde{n}_j|}
  e^{-i(\alpha_i\!-\!\tilde\alpha_j)})(1\!-\!r^{-1}x^{\frac{1}{2}\!+\!|n_i\!-\!\tilde{n}_j|}
  e^{-i(\alpha_i\!-\!\tilde\alpha_j)})}\ ,
\end{equation}
where $r=\sqrt{\frac{y_1}{y_2}}$ is for the Cartan of one factor of $SU(2)$ in
$SO(4)\!=\!SU(2)\!\times\!SU(2)$. Another combination $\sqrt{y_1y_2}$ of $U(1)^2$
chemical potentials does not appear for the ground states, as the ground states are
neutral under it. Let us provide more explanations on this expression. $\epsilon_0=\sum_{i,j=1}^N|n_i\!-\!\tilde{n}_j|\!-\!\sum_{i<j}|n_i\!-\!n_j|
\!-\!\sum_{i<j}|\tilde{n}_i\!-\!\tilde{n}_j|$ is the `zero point energy,' or more
precisely $\epsilon+j$, of monopole operators. The symmetry factor on the right hand
side is determined by the Weyl group of the subgroup of $U(N)\times U(N)$ unbroken by
the monopole charges: see \cite{Kim:2009wb} for the details. The $2N$ variables
$\alpha_i,\tilde\alpha_i$ are all integrated from $0$ to $2\pi$.
The $2N^2$ factors in the denominators of the last factor
come from exciting two anti-bifundamental scalars
in the $N\times N$ matrices. Finally, let us explain how to understand (\ref{index})
as an expression counting the ground states. The last product expression
on the right hand side comes from the multi-particle exponential (or the so-called `Plethystic'
exponential) of the `letter index.' We have reduced the general letter indices to
those carrying minimal number of $x$ factors, which suffices for studying ground states.
However, after carrying out the integration in (\ref{index}), there still appear various
terms with different powers of $x$. The expression (\ref{index}) is to be understood as
the collection of terms with minimal power of $x$. See the treatments below in this section
for some details.

We mainly study the ground state degeneracy of monopoles with $U(2)\times U(2)$
gauge group. In this case, the monopole charges are given by $H=(n_1,n_2)$ and
$\tilde{H}=(\tilde{n}_1,\tilde{n}_2)$, which are subject to
$n_1\!+\!n_2\!=\!\tilde{n}_1\!+\!\tilde{n}_2$. The last condition comes from  the fact that
an overall $U(1)$ in $U(N)\times U(N)$ decouples from the matters, constraining
a component of the magnetic field to be zero via Gauss' law \cite{Aharony:2008ug}.
For simplicity, we consider the case in which all fluxes are nonnegative.
The nature of the degeneracy depends on whether any two integers in this entry are equal
or not. It turns out that the study can be divided into three different cases as we explain now.

In the `generic' case, all four integers in $H$, $\tilde{H}$ are different.
It is sufficient to study the case with $n_1\!>\!\tilde{n}_1\!>\!\tilde{n}_2\!>\!n_2$,
as other cases with different orderings can be studied similarly.
The lowest energy state can be obtained as follows. The phase
\begin{equation}\label{phase}
  e^{ik\sum_{i=1}^N(n_i\alpha_i\!-\!\tilde{n}_i\tilde\alpha_i)}
\end{equation}
in the integrand of (\ref{index}) has to be canceled by the phases in the last factor
in (\ref{index}), after geometrically expanding the denominator, to have nonzero term
after integration. In the last factor, every factor of phase is associated with an
energy cost, i.e. comes with a positive power of $x$. To get the ground state index,
we should consider cases minimizing this energy cost. Let us temporarily ignore the
factors in the numerator, to simplify the discussion, which shall be restored shortly.
It turns out that, to minimize the energy, one should first take
$kn_2$ factors of $e^{-i(\alpha_2\!-\!\tilde\alpha_2)}$ to cancel the $e^{ikn_2\alpha_2}$
phase in (\ref{phase}). This is because the phase $e^{-i(\alpha_2\!-\!\tilde\alpha_2)}$
is associated with an energy cost $\frac{1}{2}\!+\!\tilde{n}_2\!-\!n_2$ (i.e. coming with
a factor of $x^{\frac{1}{2}\!+\!\tilde{n}_2\!-\!n_2}$), which is smaller
than the cost $\frac{1}{2}\!+\!\tilde{n}_1\!-\!n_2$ that is caused by using the phase
$e^{-i(\alpha_2\!-\!\tilde\alpha_1)}$. Similarly, all $e^{-ik\tilde{n}_1\tilde\alpha_1}$
phases should be canceled by taking $k\tilde{n_1}$ factors of
$e^{-i(\alpha_1\!-\!\tilde\alpha_1)}$ since it has lower energy cost than using the phase
$e^{-i(\alpha_2\!-\!\tilde\alpha_1)}$. One is therefore left with yet uncanceled phase
\begin{equation}
  e^{ik(n_1\!-\!\tilde{n}_1)\alpha_1+ik(\tilde{n}_2\!-\!n_2)\tilde\alpha_2}=
  e^{ik(n_1\!-\!\tilde{n}_1)(\alpha_1\!-\!\tilde\alpha_2)}\ ,
\end{equation}
where the last expression is obtained using $n_1\!+\!n_2\!=\!\tilde{n}_1\!+\!\tilde{n}_2$.
This phase should be canceled by taking $k(n_1\!-\!\tilde{n}_1)$ factors of
$e^{-i(\alpha_1\!-\!\tilde\alpha_2)}$ phases. The coefficient of
$e^{-in(\alpha_i\!-\!\tilde\alpha_j)}$ after expanding the denominator is
\begin{equation}\label{character}
  \chi_{n}(r)\equiv\frac{r^{n\!+\!1}-r^{-(n\!+\!1)}}{r-r^{-1}}=
  r^n+r^{n\!-\!2}+\cdots+r^{-n}\ ,
\end{equation}
which is the $SU(2)$ character for a representation with dimension $n\!+\!1$. Of course
the $SU(2)$ here is part of the $SO(6)$ R-symmetry. From the
three group of phases above, one obtains
\begin{equation}\label{1st-character}
  \chi_{kn_2}\chi_{k\tilde{n}_1}\chi_{k(n_1\!-\!\tilde{n}_1)}
\end{equation}
after integration. The associated factor of $x$ is given by
\begin{equation}\label{generic-energy}
  x^{\epsilon_0}x^{\frac{k}{2}(n_1\!+\!n_2)+k(\tilde{n}_2\!-\!n_2)n_2+k(n_1\!-\!\tilde{n}_1)
  \tilde{n}_1+k(n_1\!-\!\tilde{n}_2)(n_1\!-\!\tilde{n}_1)}=x^{\epsilon_0}
  x^{\frac{k}{2}(\tilde{n}_1\!+\!\tilde{n}_2)+2k\tilde{n}_1(\tilde{n}_2\!-\!n_2)}\ ,
\end{equation}
where we eliminated $n_1$ from the last expression. The exponent
\begin{equation}
  \epsilon_0+\frac{k}{2}(\tilde{n}_1+\tilde{n}_2)+2k\tilde{n}_1(\tilde{n}_2-n_2)
\end{equation}
is the value of $\epsilon\!+\!j\!=\!q\!+\!2j$ for the ground states with given monopole
charges. In classical considerations in the large $k$ regime, we will consider the
leading $\mathcal{O}(k)$ terms.
In general, $q\!+\!2j$ is all we can measure from the index. However,
for the ground states, we can deduce the values of $q$ and $j$ separately by checking
how many scalar letters are excited. By tracing back the integral that we have
just done, we find
\begin{equation}\label{charge-index}
  q=\frac{k}{2}(\tilde{n}_1+\tilde{n}_2)\ ,\ \ j=k\tilde{n}_1(\tilde{n}_2-n_2)\ .
\end{equation}
These charges will be reproduced from the classical solutions that we find
in the next section.

Had there been no numerator factors in the last factor of (\ref{index}), the expression
(\ref{1st-character}) would have been the degeneracy of the ground states.
However, if we expand the following expression
\begin{equation}
  (1-x^{n_1-n_2}e^{-i(\alpha_1-\alpha_2)})(1-x^{\tilde{n}_1-\tilde{n}_2}
  e^{-i(\tilde\alpha_1-\tilde\alpha_2)})\times c.c.=1+({\rm positive\ powers\ of}\ x)
\end{equation}
in the numerator, what we have computed in the previous paragraph is
the contribution from the leading term $1$. We now study other terms from the numerator.
Other terms all come with extra energy cost and also changes the number of some phases
in (\ref{phase}). For such terms to contribute to the ground state,
the phase changes should somehow provide an energy gain to compensate the
energy cost. To start with, let us consider the term
\begin{equation}\label{adjoint-cost}
  -x^{\tilde{n}_1-\tilde{n}_2}e^{-i(\tilde\alpha_1-\tilde\alpha_2)}
\end{equation}
in the numerator. Compared to the analysis of the previous paragraph,
the change of the phases is such that the number of required
$e^{-i(\alpha_1-\tilde\alpha_1)}$ phase is increased by $1$, while the number of
required $e^{-i(\alpha_1-\tilde\alpha_2)}$ phase is decreased by $1$. Since the former
phase comes with lower energy cost than the latter, there is an energy gain by
$\tilde{n}_1\!-\!\tilde{n}_2$, exactly canceling the cost shown in (\ref{adjoint-cost}).
One can easily see that other terms in the numerator are ignorable as they always have
net energy costs than (\ref{generic-energy}). So collecting all,
one obtains
\begin{equation}\label{net-character}
  \chi_{kn_2}\chi_{k\tilde{n}_1}\chi_{k(n_1\!-\!\tilde{n}_1)}-
  \chi_{kn_2}\chi_{k\tilde{n}_1\!+\!1}\chi_{k(n_1\!-\!\tilde{n}_1)\!-\!1}=
  \chi_{kn_2}\chi_{k(2\tilde{n}_1\!-\!n_1)}\ ,
\end{equation}
where we used the following identity for the $SU(2)$ character:
\begin{equation}\label{character-identity}
  \chi_m\chi_n-\chi_{m\!+\!1}\chi_{n\!-\!1}=\chi_{m\!-\!n}\ .
\end{equation}
(\ref{net-character}) is the final degeneracy formula for the ground states
with `generic' monopole charges satisfying
$n_1\!>\!\tilde{n}_1\!>\!\tilde{n}_2\!>\!n_2$.
Strictly speaking, (\ref{net-character}) is an index and not the true
degeneracy. But we suspect that the result (\ref{net-character}) for the
ground states could be the true degeneracy. A naive reasoning might be that, when
obtaining the expression (\ref{index}) from the general expression in \cite{Kim:2009wb},
we have truncated all contributions from matter fermions due to the requirement that
only lowest power terms in $x$ are kept in the letter indices. (Fermions carry larger
scale dimensions than scalars.) Also, in all cases that we can check with the
semiclassical solutions within the range of our ansatz, we find that the degeneracy is
either exactly the same as (\ref{net-character}) or smaller (implying that there are more
general solution than what we find), but never larger than (\ref{net-character}).

Although the intermediate steps of our derivations above
sometimes used $n_1\gneq\tilde{n}_1>\tilde{n}_2\gneq n_2$, one can separately check that
the final result (\ref{net-character}) actually holds for the case with
$n_1=\tilde{n}_1>\tilde{n}_2=n_2$ as well. The last case has been studied
in detail in \cite{Kim:2009ia}, together with a direct counting of the
degeneracy (\ref{net-character}) by quantizing classical solutions.

We also comment that, while obtaining the above index for the generic $U(2)\times U(2)$
monopole charges, the $11,21,22$ matrix elements of the matters (= scalars) are
all excited to saturate the Gauss' law (from the fact that phases $e^{-i(\alpha_1\!-\!\tilde\alpha_1)}$,
$e^{-i(\alpha_1\!-\!\tilde\alpha_2)}$, $e^{-i(\alpha_2\!-\!\tilde\alpha_2)}$ are used in the last
factor of (\ref{index}) to obtain (\ref{net-character})), while the $12$ component charged under
$\tilde{n}_1\!-\!n_2$ is not excited. This is due to the requirement that we only count
the ground states with given monopole charges. Later, when we consider
the semiclassical solutions with generic charges in section 3.2, it should be
remembered that the last component of the scalar should be turned off to
obtain the lowest energy solutions.

Now we proceed to consider other `non-generic' fluxes in $U(2)\times U(2)$ monopoles.
The second case comes with one of the two pairs of fluxes in $H$ or $\tilde{H}$ being equal.
Again without losing generality, we can restrict our study to the case with
$n_1>\tilde{n}_1\!=\!\tilde{n}_2>n_2$. We start by noting that, since the flux does not
break the second $U(2)$ gauge group, there is a degeneracy between the energy costs
in taking the two phases $e^{-i(\alpha_i-\tilde\alpha_1)}$ and $e^{-i(\alpha_i-\tilde\alpha_2)}$
from the denominator of the matter part in (\ref{index}). Again, we ignore the numerators in the
integrand for a while. Calling
$\tilde{n}\equiv\tilde{n}_1=\tilde{n}_2$, the phase (\ref{phase}) can be decomposed as
\begin{equation}\label{phase-degenerate}
  e^{ikn_1\alpha_1\!+\!ikn_2\alpha_2\!-\!ik\tilde{n}(\tilde\alpha_1\!+\!\tilde\alpha_2)}
  =\left(e^{i(\alpha_2-\tilde\alpha_1)}\right)^p
  \left(e^{i(\alpha_2-\tilde\alpha_2)}\right)^{kn_2-p}
  \left(e^{i(\alpha_1-\tilde\alpha_1)}\right)^{k\tilde{n}-p}
  \left(e^{i(\alpha_1-\tilde\alpha_2)}\right)^{k(n_1-\tilde{n})+p}
\end{equation}
for $0\leq p\leq kn_2$, and each of the four phases can be canceled by taking appropriate
numbers of phases from the matters. The resulting index is given by
\begin{equation}
  \sum_{p=0}^{kn_2}\chi_p\chi_{kn_2-p}\chi_{k\tilde{n}-p}\chi_{k(n_1\!-\!\tilde{n})+p}.
\end{equation}
To this expression, we should subtract various contributions coming from the numerator,
which is given by
\begin{equation}\label{U(2)-numerator}
  (1-x^{n_1\!-\!n_2}e^{-i(\alpha_1-\alpha_2)})(c.c.)\left(1-\frac{1}{2}
  e^{-i(\tilde\alpha_1\!-\!\tilde\alpha_2)}-\frac{1}{2}e^{-i(\tilde\alpha_2
  \!-\!\tilde\alpha_1)}\right)\ .
\end{equation}
In the last factor, we multiplied the symmetry factor $\frac{1}{2}$ in
(\ref{index}).\footnote{The last factor in (\ref{U(2)-numerator}) is the Haar measure
for the unbroken $U(2)$ gauge group.}
The last factor can be effectively replaced by
$1-e^{-i(\tilde\alpha_2\!-\!\tilde\alpha_1)}$ in the integral. One can show that the energy
cost associated with the first two factors can never be compensated by an energy gain.
Therefore, we only need
to consider the third factor. Considering the second term in this factor,
i.e. after multiplying the phase $-e^{-i(\tilde\alpha_2\!-\!\tilde\alpha_1)}$ to
the remaining part of the integrand, the total phase to be cancel by
matters can be written as
\begin{equation}\label{phase-degenerate-subtract}
  -e^{-i(\tilde\alpha_2\!-\!\tilde\alpha_1)}
  e^{ikn_1\alpha_1\!+\!ikn_2\alpha_2\!-\!ik\tilde{n}(\tilde\alpha_1\!+\!\tilde\alpha_2)}
  =\left(e^{i(\alpha_2-\tilde\alpha_1)}\right)^p
  \left(e^{i(\alpha_2-\tilde\alpha_2)}\right)^{kn_2-p}
  \left(e^{i(\alpha_1-\tilde\alpha_1)}\right)^{k\tilde{n}\!-\!1\!-\!p}
  \left(e^{i(\alpha_1-\tilde\alpha_2)}\right)^{k(n_1-\tilde{n})\!+\!1\!+\!p}
\end{equation}
with $0\leq p\leq kn_2$, from which one obtains
\begin{equation}
  -\sum_{p=0}^{kn_2}\chi_p\chi_{kn_2-p}\chi_{k\tilde{n}\!-\!1\!-\!p}
  \chi_{k(n_1\!-\!\tilde{n})\!+\!1\!+\!p}\ .
\end{equation}
Combining all, one obtains
\begin{equation}
  \sum_{p=0}^{kn_2}\chi_p\chi_{kn_2-p}\left(\chi_{k\tilde{n}-p}
  \chi_{k(n_1\!-\!\tilde{n})+p}-\chi_{k\tilde{n}\!-\!1\!-\!p}
  \chi_{k(n_1\!-\!\tilde{n})\!+\!1\!+\!p}\right)=
  \sum_{p=0}^{kn_2}\chi_p\chi_{kn_2\!-\!p}\chi_{2p\!-\!kn_2}\ ,
\end{equation}
where we used the identity (\ref{character-identity}).
Note that the last `character' $\chi_{2p-kn_2}$ can come with negative argument,
in which case the group theoretic interpretation becomes vague. However, the identity
(\ref{character-identity}) still holds with negative arguments with the definition
of $\chi_n$ given by the first form in (\ref{character}). To manipulate the last
expression, note that the summation variable $p$ can be changed to $p^\prime=kn_2\!-\!p$.
Averaging over the two identical expression, and using
$\chi_n(r)+\chi_{-n}(r)=r^n+r^{-n}$, one obtains
\begin{equation}
  \frac{1}{2}\sum_{p=0}^{kn_2}\chi_p\chi_{kn_2\!-\!p}\left(r^{2p\!-\!kn_2}+
  r^{-2p\!+\!kn_2}\right)\ .
\end{equation}
After expanding the remaining characters and appropriately reorganizing, one obtains
\begin{equation}
  {\rm index}=\left\{
  \begin{array}{ll}\chi_{kn_2}^2+\chi_{kn_2-2}^2+\cdots+\chi_0^2&{\rm for\ even}\ kn_2\\
  \chi_{kn_2}^2+\chi_{kn_2-2}^2+\cdots+\chi_1^2&{\rm for\ odd}\ kn_2\ .
  \end{array}\right.
\end{equation}
We see that the first term on each line is a reduction of the index (\ref{net-character})
in the generic case. Therefore, with $\tilde{n}_1\!=\!\tilde{n}_2$, there appears extra
sectors in the ground states. We shall briefly discuss this point in the next section
with our classical solutions.

Finally, the fluxes can come with $n_1\!=\!n_2\!=\!\tilde{n}_1\!=\!\tilde{n}_2$.
Calling this integer $n$,
one can show that the resulting index is
\begin{equation}
  \frac{1}{2}\chi_{kn}(r^2)+\frac{1}{2}\chi_{kn}(r)^2\ .
\end{equation}
This result is also obtained by quantizing classical solutions \cite{Kim:2009ia},
as the fluxes satisfy $H=\tilde{H}$.

To summarize, the $U(2)\times U(2)$ index is given by
\begin{equation}\label{U(2)-index}
  I_{(n_1,n_2)(\tilde{n}_1,\tilde{n}_2)}=\left\{
  \begin{array}{ll}\chi_{kn_2}(r)\chi_{k(2\tilde{n}_1\!-\!n_1)}(r)&
  {\rm for}\ \ n_1\!\geq\!\tilde{n}_1\!>\!\tilde{n}_2\!\geq\!n_2\\
  \chi_{kn_2}^2+\chi_{kn_2-2}^2+\cdots+\chi_0^2&{\rm for}\ \
  n_1\!>\!\tilde{n}_1\!=\!\tilde{n}_2\!>\!n_2\ \ {\rm and\ even}\ kn_2\\
  \chi_{kn_2}^2+\chi_{kn_2-2}^2+\cdots+\chi_1^2&{\rm for}\ \
  n_1\!>\!\tilde{n}_1\!=\!\tilde{n}_2\!>\!n_2\ \ {\rm and\ odd}\ kn_2\\
  \frac{1}{2}\chi_{kn}(r^2)+\frac{1}{2}\chi_{kn}(r)^2&{\rm for}\ \
  n_1\!=\!n_2\!=\!\tilde{n}_1\!=\!\tilde{n}_2\!\equiv\!n
  \end{array}
  \right.\ .
\end{equation}
It is helpful to consider the simpler case with $n_2=0$, as this case will turn out to
be very simple from the semi-classical analysis. One finds
\begin{equation}\label{U(2)-index-n2=0}
  I_{(\tilde{n}_1+\tilde{n}_2,0),(\tilde{n}_1,\tilde{n}_2)}=
  \chi_{k(\tilde{n}_1\!-\!\tilde{n}_2)}(r)\ ,
\end{equation}
which alludes to a contribution from one irreducible representation of global
$SU(2)$. We will show in section 3.1 that this degeneracy indeed comes from
quantizing the classical moduli of the monopole solutions, generated by the
global $SU(2)$ zero modes.

Although the $U(2)\times U(2)$ monopoles are the main subject of this paper,
one might wonder how the above degeneracy formula for the ground states would
generalize for $U(N)\times U(N)$ monopoles. As far as we can see, the generalization
is not so straightforward and exhibits new features. For instance, we find that
the ground state index for the $U(3)\times U(3)$ monopoles depends
sensitively on the order of the monopole charges $H=(n_1,n_2,n_3)$,
$\tilde{H}=(\tilde{n}_1,\tilde{n}_2,\tilde{n}_3)$. We simply record one
peculiar behavior that we find for certain monopole charges. Namely, when the
monopole charges satisfy the following conditions,
\begin{equation}
  n_1\!>\!\tilde{n}_1\!>\!\tilde{n}_2\!>\!\tilde{n}_3\!>\!n_2\!>\!
  n_3\ \ {\rm with}\ \ \tilde{n}_3\!>\!n_2\!+\!n_3\ ,
\end{equation}
we find that the index is zero for all $k$. It would be quite curious to see
if this implies that there are simply no states at this energy.

\section{Classical monopole solutions}

In this section, we consider classical solutions for the monopoles
in the $\mathcal{N}\!=\!6$ Chern-Simons-matter theory preserving
some supersymmetry. After explaining the basic setting, involving
the details of the field theory on $S^2\times\mathbb{R}$, we explain
the general equations for the BPS monopoles that we are interested in.
The details of the solutions for various monopole charges are considered
in the following subsections.

To study local operators preserving minimal number of supersymmetry,
one starts by picking a complex supercharge among 12 Poincare
supercharges $Q_{IJ\alpha}\!=\!-Q_{JI\alpha}$ (where $I,J=1,2,3,4$, $\alpha=\pm$). Without
losing generality, we pick $Q_{34-}$ as in \cite{Bhattacharya:2008bja,Kim:2009wb}.
After radial quantization, to be explained in detail below, one can regard
the conformal supercharges $S^{IJ\alpha}$ as Hermitian conjugates of
$Q_{IJ\alpha}$. The local operators, or states, that we are interested in
are annihilated by $Q\!=\!Q_{34-}$ and $S\!=\!S^{34-}$. From the algebra
\begin{equation}
  \{Q,S\}\sim\epsilon-q-j\ ,
\end{equation}
operators (states) annihilated by $Q,S$ satisfies the BPS bound $\epsilon\!=\!q\!+\!j$.
The Poincare supercharges $Q_\alpha\equiv Q^{34}_\alpha$ define an $\mathcal{N}\!=\!2$
subgroup of the full $\mathcal{N}\!=\!6$ supersymmetry. It will be convenient
to employ the $\mathcal{N}\!=\!2$ superfield notations as in
\cite{Klebanov:2008vq,Kim:2009wb}, which we do from now on.

In the $\mathcal{N}\!=\!2$ supersymmetric formulation, the matter fields can be
decomposed to two bifundamental
chiral supermultiplets $\hat\phi_a,\psi_{a\alpha}$ and two anti-bifundamental
chiral supermultiplets $\phi_a,\chi_{a\alpha}$ ($a=1,2$). In addition, there are
$U(N)\times U(N)$ vector multiplet $A_\mu,\sigma,\lambda_\alpha$ and
$\tilde{A}_\alpha,\tilde\sigma,\tilde\lambda_\alpha$. The adjoint scalars
$\sigma,\tilde\sigma$ and fermions $\lambda_\alpha,\tilde\lambda_\alpha$ are composite fields
which are quadratic in the matter fields. In particular,
\begin{equation}
  \sigma=\frac{2\pi}{k}\left(\hat\phi_a\hat\phi_a^\dag-\phi_a^\dag\phi_a\right)\ ,\ \
  \tilde\sigma=\frac{2\pi}{k}\left(\hat\phi_a^\dag\hat\phi_a-\phi_a\phi_a^\dag\right)\ .
\end{equation}
The action and supersymmetry
transformation in our notation can be found in, say, \cite{Kim:2009wb}. An aspect
worth an explanation is the so-called `baryon-like' $U(1)_b$ charge, which is the
global part of the local $U(1)$ gauge transformation associated with
${\rm tr}A_\mu-{\rm tr}\tilde{A}_\mu$. The current of this symmetry is related by
Gauss' law
\begin{equation}
  j_\mu^{U(1)_b}={\rm tr}\left[i\left(\phi_aD_\mu\phi_a^\dag-D_\mu\phi_a\phi_a^\dag\right)
  +i\left(\hat\phi_a^\dag D_\mu\hat\phi_a-D_\mu\hat\phi_a^\dag\hat\phi_a\right)\right]
  =\frac{k}{2\pi}{\rm tr}\left(\star F_\mu\right)=\frac{k}{2\pi}{\rm tr}
  \left(\star \tilde{F}_\mu\right)
\end{equation}
to the magnetic flux of
$\frac{k}{2\pi}{\rm tr}F_{\mu\nu}\!=\!\frac{k}{2\pi}{\rm tr}\tilde{F}_{\mu\nu}$
on $S^2$. Our convention is such that the last magnetic flux is positive
when the fields $\phi_a^\dag$ and $\hat\phi_a$ are excited.

Now we explain the radially quantized CFT.
Conformal field theories can be defined in arbitrary conformally
flat background. In particular, it is often useful to consider
such theories on a spatial round sphere. The motivation for this setting
is that $S^{d\!-\!1}\times\mathbb{R}$ is the boundary of
global $AdS_{d\!+\!1}$ spacetime. We consider the
Chern-Simons-matter theory living on a 2-sphere with unit radius.
The metric on \textit{Minkowskian} $S^2\times\mathbb{R}$ is
\begin{equation}\label{metric}
  ds^2=-dt^2+d\theta^2+\sin^2\theta d\varphi^2\ ,
\end{equation}
where the overall radius is set to $1$.
A simple way of obtaining a CFT living on the background (\ref{metric})
from the CFT on $\mathbb{R}^{2+1}$ is called the radial quantization.
We first consider a CFT on Euclidean $\mathbb{R}^3$, and relate the
radial variable $r$ of $\mathbb{R}^3$ (around any point) and the
time $\tau$ of the CFT living on \textit{Euclidean} $S^2\times\mathbb{R}$
as $r=e^{\tau}$. The fields in the two Euclidean theories are related
as follows. Firstly, abstractly denoting by $\phi,\phi_S$ the scalars
in the former and latter theories, the two are related by
\begin{equation}
  \phi=r^{-\frac{1}{2}}\phi_S\ ,
\end{equation}
where the exponent $\frac{1}{2}$ comes from the dimension of the scalar $\phi$
on $\mathbb{R}^3$. Also, the gauge field $A_\mu$ as a 1-form on $\mathbb{R}^3$
is simply taken to be the 1-form on $S^2\times\mathbb{R}$
\begin{equation}
  A=A_rdr+A_\theta d\theta+A_\phi d\varphi=
  A_\tau d\tau+A_\theta d\theta+A_\varphi d\varphi
\end{equation}
with $rA_r=A_\tau$ understood from the coordinate transformation. The relation
between the fermionic fields of the two theories can be found in the appendix of
\cite{Kim:2009wb}. Plugging all these field transformations into the action of
the CFT on $\mathbb{R}^3$, one obtains a CFT action on Euclidean
$S^2\times\mathbb{R}$. Finally, a continuation $\tau=it$ (with $A_t=iA_\tau$)
yields a CFT on Minkowskian $S^2\times\mathbb{R}$, where $\mathbb{R}$
is generated by $t$. An important feature of the last theory is that the scalars
acquire conformal mass terms with masses given by $m^2=\frac{1}{4}$.

Local operators inserted at a point
($r=0$ in our explanation) are in 1-to-1 correspondence with the
states in the radially quantized theory. The scale dimension of operators map to
the energy of the corresponding states. In particular, monopole operators create
magnetic flux on spatial $S^2$ so that we are lead to study states propagating on
$S^2\times\mathbb{R}$ in the presence of magnetic fields.

Having the above field and coordinate transformations in mind, we shall freely go
back and forth between the expressions on $\mathbb{R}^3$ and $S^2\times \mathbb{R}$ in
our analysis below. Since we always take $t=-i\tau$ to be real,
$r\!=\!e^{\tau}\!=\!e^{it}$ should be regarded as a phase in all analysis on
$\mathbb{R}^3$. Accordingly, the Cartesian coordinates $x^\mu$ ($\mu\!=\!1,2,3$)
of $\mathbb{R}^3$ are not real but are subject to the following complex conjugation
rule: $(x^\mu)^\ast=\frac{x^\mu}{r^2}$.

For simplicity, we consider the monopoles with all
integer fluxes $n_i$, $\tilde{n}_i$ in (\ref{monopole-charge})
being positive. With all fluxes being positive, the lowest energy states
(or the operators with lowest scale dimension) for given
monopole charges come in the sector in which only the gauge fields and the
anti-bifundamental scalars $\phi_a$ are excited. The Gauss' law is given by
\begin{equation}\label{gauss-general}
  \frac{k}{2\pi}\star F_\mu=i\left(D_\mu\phi_a^\dag\phi_a-\phi_a^\dag D_\mu\phi_a\right)\ ,\ \
  \frac{k}{2\pi}\star\tilde{F}_\mu=i\left(\phi_a D_\mu\phi_a^\dag-
  D_\mu\phi_a\phi_a^\dag\right)\ .
\end{equation}
This expression holds for the theory defined either on $\mathbb{R}^{2+1}$
or Minkowskian $S^2\times\mathbb{R}$, where the Hodge dual $\star$ is taken
with appropriate metric for each case. The supersymmetry conditions
$Q_-\chi_{a\alpha}=0$ (on $\mathbb{R}^3$) for spin indices $\alpha=\pm$ are given by
\begin{equation}\label{susy-general}
  (D_1-iD_2)\phi_a^\dag=0\ ,\ \
  D_3\phi_a^\dag+(\sigma\phi_a^\dag-\phi_a^\dag\tilde\sigma)=0\ ,
\end{equation}
where the subscripts $1,2,3$ are for the three Cartesian coordinates of
$\mathbb{R}^3$. $x^3$ coordinate is chosen so that $\alpha=\pm$ components
for spinors come with $j\!\equiv\!\pm\frac{1}{2}$. One can check that
these supersymmetry conditions imply the following equation of motion
(see \cite{Kim:2009wb}),
\begin{equation}\label{eom}
  D^\mu D_\mu\phi_a^\dag+(\sigma\phi_a^\dag-\phi_a^\dag\tilde\sigma)\tilde\sigma-
  \sigma(\sigma\phi_a^\dag-\phi_a^\dag\tilde\sigma)+\phi_a^\dag\tilde{D}-D\phi_a^\dag=0\ ,
\end{equation}
where the last two terms come from the on-shell values of $D$-term fields
as one uses the equation of motion for $\sigma,\tilde\sigma$.
To see this, we consider
\begin{eqnarray}\label{eom-step}
  0&=&(D_1+iD_2)(D_1-iD_2)\phi_a^\dag+D_3\left[D_3\phi_a^\dag
  +(\sigma\phi_a^\dag-\phi_a^\dag\tilde\sigma)\right]\nonumber\\
  &=&D^\mu D_\mu\phi_a^\dag-i[D_1,D_2]\phi_a^\dag
  +D_3(\sigma\phi_a^\dag-\phi_a^\dag\tilde\sigma)\ .
\end{eqnarray}
Expanding the last two terms on the second line, one obtains
\begin{equation}
  -F_{12}\phi_a^\dag+\phi_a^\dag\tilde{F}_{12}+D_3\sigma\phi_a^\dag-\phi_a^\dag
  D_3\tilde\sigma-\sigma(\sigma\phi_a^\dag-\phi_a^\dag\tilde\sigma)+
  (\sigma\phi_a^\dag-\phi_a^\dag\tilde\sigma)\tilde\sigma\ .
\end{equation}
We combine the first four terms using $F_{12}-D_3\sigma=D$ and
$\tilde{F}_{12}-D_3\tilde\sigma=\tilde{D}$, which come from the supersymmetry
conditions $Q_-\lambda_-=0$ and $Q_-\tilde\lambda_-=0$ for the gaugino composites
\cite{Kim:2009wb}, to obtain $\phi_a^\dag\tilde{D}-D\phi_a^\dag$. Then (\ref{eom-step})
reduces to the equation of motion (\ref{eom}), proving the assertion.

In the following two subsections, we solve and discuss the Gauss' law condition
(\ref{gauss-general}) and the supersymmetry condition (\ref{susy-general}) with
various $U(2)\times U(2)$ monopole charges.

\subsection{General $U(1)\times U(2)$ monopoles}

We first consider the monopoles with magnetic charges given by
$H=(n_1\!+\!n_2,0)$ and $\tilde{H}=(n_1,n_2)$: namely, one of the
$U(1)^2$ charges in the first gauge group is taken to be zero. (We take $n_1,n_2>0$.)
In fact, the field contents that we excite in this solution stays
within the $U(1)\times U(2)$ Chern-Simons-matter theory. The solutions
in this case will turn out to be significantly simpler than the monopoles
in $U(2)\times U(2)$ with general magnetic flux $H=(n_1,n_2)$,
$\tilde{H}=(\tilde{n}_1,\tilde{n}_2)$, whose study is postponed to
the next subsection.

Since nonzero fluxes are turned on in the diagonal of $U(2)\times U(2)$
in certain basis, we start from the following ansatz
\begin{equation}\label{n2=0-gauge}
  A_\mu=\left(\begin{array}{cc}A^1_\mu&0\\0&0\end{array}\right)\ ,\ \
  \tilde{A}_\mu=\left(\begin{array}{cc}\tilde{A}^1_\mu&0\\0&\tilde{A}_\mu^2\end{array}
  \right)
\end{equation}
for the gauge fields. The electric component $\mu\!=\!t$ of the gauge field will turn out
to be necessary for this configuration to satisfy all supersymmetry conditions and Gauss' law,
while we consistently set $\mu\!=\!\theta$ components to zero. All nonzero components
are taken to depend on $\theta$ coordinates only. To have the Gauss' law compatible with
this form of gauge fields, it is easy to check that the following form of the
anti-bifundamental scalars
\begin{equation}\label{n2=0-scalar}
  \phi_1=\left(\begin{array}{cc}\psi e^{im_1\varphi-i\omega_1 t}&0\\0&0\end{array}\right)
  \ ,\ \ \phi_2=\left(\begin{array}{cc}0&0\\ \chi e^{im_2\varphi-i\omega_2 t}&0
  \end{array}\right)\ ,
\end{equation}
provide consistent ansatz,
where $\psi,\chi$ are complex functions of $\theta$. The fact that second columns of scalars
are all zero implies that the scalars can be regarded as $U(1)\times U(2)$ anti-bifundamental
fields. Of course, this ansatz can be generalized by using the $SU(2)_R$ global symmetry
which rotates $\phi_1,\phi_2$ as a doublet. In the notation of our previous section, the Cartan
of this $SU(2)_R$ is conjugate to the chemical potential $r$. This will generate a
constant moduli of the solution, which we shall describe in detail later. We find that the
$SU(2)_R$ action on (\ref{n2=0-scalar}) is the most general form compatible with
(\ref{n2=0-gauge}) and the Gauss' law. We also note that, using appropriate
local $U(1)^2$ gauge transformations in the second $U(2)$ gauge group, one can
eliminate the $\varphi,t$ dependent phases in the scalars by shifting
$\tilde{A}_\mu^1$ and $\tilde{A}_\mu^2$ by suitable constants. Below, we
assume this form of scalars with $\varphi,t$ dependent phases eliminated.

Plugging the above ansatz into the supersymmetry conditions and rearranging,
one obtains the following differential and algebraic conditions
\begin{eqnarray}\label{susy-n2=0}
  x(1-x^2)f_{1,2}^\prime&=&-2\left[(1-x^2)\left(g_{1,2}+\frac{1}{2}\right)+h_{1,2}
  \right]f_{1,2}\nonumber\\
  xf_{1,2}&=&-\left(g_{2,1}+h_{2,1}+\frac{1}{2}\right)
\end{eqnarray}
where $x\equiv\cos\theta$. Prime denotes $x$ derivative, and
$f_1\equiv\frac{2\pi}{k}|\psi|^2$, $f_2\equiv\frac{2\pi}{k}|\chi|^2$,
$g_{1,2}\equiv A^1_t-\tilde{A}^{1,2}_t$, $h_{1,2}\equiv A^1_\varphi-\tilde{A}^{1,2}_\varphi$.
Also, the phases of $\psi$, $\chi$ are required to be constants (i.e. $\theta$ independent)
from the supersymmetry conditions. Note that the components of the gauge fields appear only
in the combinations of $g_{1,2}$, $h_{1,2}$, as the matter fields are neutral under
overall $U(1)$ of $U(2)\times U(2)$.

With our ansatz, the Gauss' law conditions reduce to
\begin{eqnarray}\label{gauss-n2=0}
  &&h_1^\prime=2g_2f_2\ ,\ \ (1-x^2)g_1^\prime=2h_2f_2\nonumber\\
  &&h_2^\prime=2g_1f_1\ ,\ \ (1-x^2)g_2^\prime=2h_1f_1\ ,
\end{eqnarray}
while $A^1_t,A^1_\varphi$ is given by
\begin{equation}
  A^1_t=\tilde{A}^1_t+\tilde{A}^2_t+const.\ ,\ \
  A^1_\varphi=\tilde{A}^1_\varphi+\tilde{A}^2_\varphi+const.
\end{equation}
from ${\rm tr}F_{\mu\nu}\!-\!{\rm tr}\tilde{F}_{\mu\nu}\!=\!0$. From the
algebraic conditions in (\ref{susy-n2=0}), the Gauss' law conditions (\ref{gauss-n2=0})
are equations for the four functions $g_1,g_2,h_1,h_2$. One can easily show that
the Gauss' law conditions (\ref{gauss-n2=0}) imply the differential condition for $f_1,f_2$
in (\ref{susy-n2=0}). Therefore, it suffices for us to solve
(\ref{gauss-n2=0}), with the algebraic conditions in (\ref{susy-n2=0}) understood.
Since we want solutions with given flux $H=(n_1\!+\!n_2,0)$, $\tilde{H}=(n_1,n_2)$,
the solutions should satisfy
\begin{eqnarray}\label{quantization-n2=0}
  n_1&=&\frac{1}{2\pi}\int_{S^2}d\theta d\varphi\ \partial_\theta \tilde{A}^1_\varphi
  =-\int_{-1}^1dx(A^1_\varphi\!-\!\tilde{A}^2_\varphi)^\prime=h_2(-1)-h_2(1)\ ,\nonumber\\
  n_2&=&h_1(-1)-h_1(1)\ ,
\end{eqnarray}
where the arguments in the functions stand for $x\!=\!\pm 1$.
After solving the differential equations (\ref{gauss-n2=0}), the quantization of
the right hand sides of (\ref{quantization-n2=0}) should be imposed by hand,
just like the quantized magnetic charge of the Dicac monopole.

Let us compute the conserved
Noether charges. Firstly, the U(1) $R$-charge $q$ appearing in the BPS bound is given by
\begin{equation}
  q=\int_{S^2}{\rm tr}\left[\frac{i}{2}D_t\phi_a\phi_a^\dag-
  \frac{i}{2}\phi_aD_t\phi_a^\dag\right]=-k\int_{-1}^1dx\left[g_1f_1+g_2f_2\right]=
  -\frac{k}{2}\int_{-1}^1dx\left[h_1^\prime+h_2^\prime\right]=\frac{k}{2}(n_1+n_2)\ ,
\end{equation}
where we used (\ref{gauss-n2=0}) at the third step.
This is consistent with (\ref{charge-index}) from the index.
Also, the angular momentum along the
$x^3$ direction of $\mathbb{R}^3$ is given by
\begin{eqnarray}
  j&=&-\int_{S^2}{\rm tr}\left[D_t\phi_aD_\varphi\phi_a^\dag
  +D_\varphi\phi_aD_t\phi_a^\dag \right]=-k\int_{-1}^1 dx
  \left[h_1(x)h_2(x)\right]^\prime\ ,
\end{eqnarray}
after using (\ref{gauss-n2=0}).
As we shall explain shortly in the construction of solutions, the functions
$h_1,h_2$ should satisfy the boundary conditions $h_1(1)=h_2(1)=0$. Thus, one obtains
\begin{equation}
  j=kn_1n_2\ ,
\end{equation}
again consistent with (\ref{charge-index}). We have also explicitly checked that
the Noether energy of the solution satisfies the BPS bound $\epsilon=q+j$.

Now we turn to find the solutions of (\ref{gauss-n2=0}). As this equation is
nonlinear, we do not know how to construct analytic solutions. We therefore
construct numerical solutions.

Before presenting our numerical solutions, we explain the boundary conditions
to be imposed for various functions at $x\!=\!\pm 1$ (i.e. $\theta\!=\!0,\pi$).
The equations containing $g_{1,2}^\prime$ in (\ref{gauss-n2=0}) demand that the
right hand sides $2h_2f_2$, $2h_1f_1$ be zero at $x\!=\!\pm 1$
to have regular solutions:
otherwise, the factors $(1\!-\!x^2)$ on the left hand sides would render the
functions $g_{1,2}$ (and thus other functions through back-reactions) to diverge
at $x\!=\!\pm 1$.
With $n_{1,2}>0$, the anti-bifundamental modes $\psi$, $\chi$ feel negative fluxes
$-n_2$ and $-n_1$, respectively. In this case, one can see that
\begin{equation}\label{boundary-n2=0}
  h_1(1)\!=\!h_2(1)\!=\!0\ ,\ \ f_1(-1)\!=\!f_2(-1)\!=\!0
\end{equation}
should be imposed to have $h_{1,2}(\pm 1)f_{1,2}(\pm 1)\!=\!0$ satisfied.
This can be motivated by recalling  the properties of the magnetic monopole
harmonics \cite{Wu:1976ge}, which are the wavefunctions of charged scalars under
uniform magnetic field.\footnote{Although our scalars $\psi,\chi$ are not given by
these monopole harmonics as the magnetic field is not uniform, the properties
of the monopole harmonics that we are going to explain are all topological so
that they should apply to our solutions as well.} We pay attention to the
`highest weight states' for which the $SO(3)$ Cartan $j$ is equal to the total
angular momentum, which is the case for the BPS states that we consider.
The wavefunctions $Y^{(N)}_{q,j,j}$ in the `north' and `south' patches are given
by
\begin{equation}
  Y^{(N)}_{qjj}=(1-x)^{\frac{j+q}{2}}(1+x)^{\frac{j-q}{2}}
  e^{i(j+q)\varphi}\ ,\ \ Y^{(S)}_{qjj}=(1-x)^{\frac{j-q}{2}}(1+x)^{\frac{j+q}{2}}
  e^{i(j-q)\varphi}\ ,
\end{equation}
where $q$ is magnetic field felt by the charged field. For $n_{1,2}>0$, we take $q$
to be negative. Since we employed the gauge in which there are no $\varphi$ dependent
phases, our solution should be presented in the `north patch gauge' in which
$j\!+\!q\!=\!0$. In our `north' gauge, the magnetic parts of the gauge fields $h_{1,2}$ should
all be zero at the north pole $x\!=\!1$. Also, the $\theta$ dependent profile of the
harmonics becomes $(1+x)^j$, which is zero at the south pole $x\!=\!-1$ but nonzero
at the north pole. Our claim (\ref{boundary-n2=0}) is demanding the same properties.

By studying the equations (\ref{gauss-n2=0}), one obtains the following asymptotic solutions
\begin{eqnarray}\label{x=1-BC}
  g_{1,2}(x)&=&g_{1,2}(1)+2g_{1,2}(1)\left(g_{1,2}(1)+\frac{1}{2}\right)
  \left(g_{2,1}(1)+\frac{1}{2}\right)(1-x)+
  \mathcal{O}(1\!-\!x)^2\nonumber\\
  h_{1,2}(x)&=&2g_{2,1}(1)\left(g_{1,2}(1)+\frac{1}{2}\right)(1-x)+\mathcal{O}(1\!-\!x)^2
\end{eqnarray}
near $x\!=\!1$, and
\begin{eqnarray}\label{x=-1-BC}
  g_{1,2}(x)&=&-\left(n_{2,1}+\frac{1}{2}\right)+a_{1,2}(1+x)^{n_{1,2}}+\cdots\nonumber\\
  h_{1,2}(x)&=&n_{2,1}-\frac{2n_{1,2}+1}{n_{1,2}+1}\ a_{1,2}(1+x)^{n_{1,2}+1}+\cdots
\end{eqnarray}
near $x\!=\!-1$. At $x\!=\!1$, we only have two free parameters $g_{1,2}(1)$
after demanding that the solution be regular there. On the other hand, we have all four
parameters $n_{1,2},a_{1,2}$ in the asymptotic solution near $x\!=\!-1$, even after demanding
regularity. The two `flux'
parameters $n_{1,2}$ are continuous when we consider differential equations,
which we shall quantize by hands later for quantum consistency.
Once we start the construction of numerical solutions by imposing two boundary conditions
$g_{1,2}(1)$ in (\ref{x=1-BC}), the solutions will flow to the desired solution of the
form (\ref{x=-1-BC}) for generic choice of $g_{1,2}(1)$ (chosen in a suitable range),
as (\ref{x=-1-BC}) has 4 parameters to fit general numerical solutions.
The boundary conditions $g_{1,2}(1)$ determine two independent fluxes $n_{1,2}$, whose
precise relation is to be found by numerics.

\begin{figure}[t]
  \begin{center}
   \includegraphics[width=8cm]{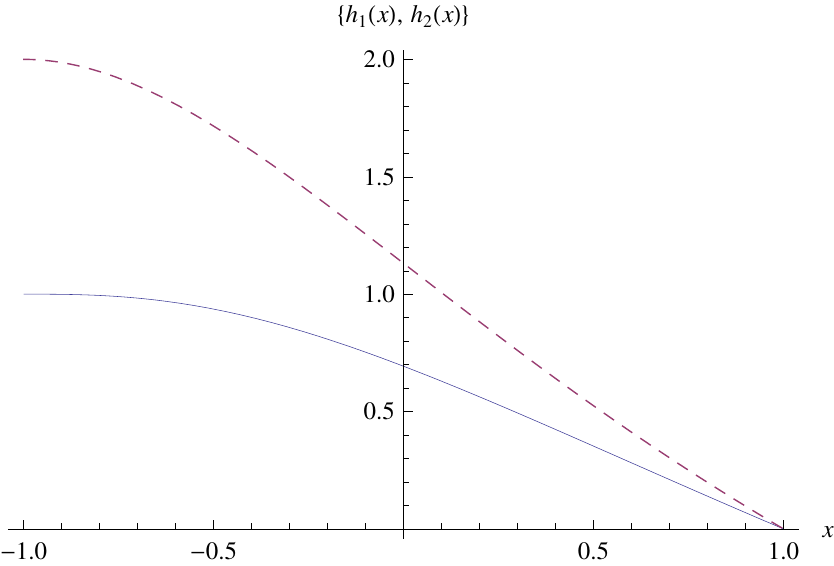}\hspace{0.5cm}
   \includegraphics[width=8cm]{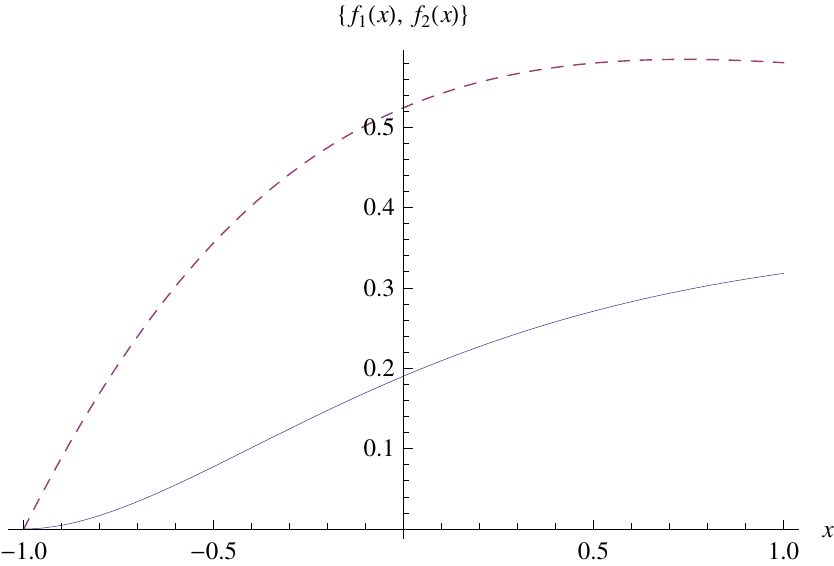}
\caption{Profiles of $h_1$ (normal line), $h_2$ (dashed) on the left, and
$f_1$ (normal), $f_2$ (dashed) on the right. The parameters $g_1(1),g_2(1)$
are tuned to have fluxes $H\!\cong\!(3,0)$, $\tilde{H}\!\cong\!(2,1)$.}\label{(30)(21)}
  \end{center}
\end{figure}
To deal with the fine-tuning of the boundary conditions at $x\!=\!1$ for regularity,
we choose $g_{1,2}(1)$ appropriately and then start the numerics from $x\!=\!.9999$
with the values of $g_{1,2}(.9999)$ and $h_{1,2}(.9999)$ determined by (\ref{x=1-BC}).
With a choice of $g_{1,2}(1)$ in suitable range, we obtain solutions which are regular
in $-1\!<\!x\!<\!1$. In Fig.\ref{(30)(21)},
we plot a numerical solution with $h_2(-1)\!=\!n_1\!=\!2.00475$ and
$h_1(-1)\!=\!n_2\!=\!1.00197$, or $H\!\cong\!(3,0)$ and $\tilde{H}\!\cong\!(2,1)$.
The parameters at $x\!=\!1$ for this configuration are chosen to be $g_1(1)\!=\!-.818$,
$g_2(1)\!=\!-1.0801$. Solutions with other monopole charges can also be found.

Now we construct (what we believe is) the most general solution with $n_2\!=\!0$, by
including the moduli generated by $SU(2)_R$ action on the ansatz (\ref{n2=0-scalar})
that we already mentioned. Acting the following $SU(2)_R$ matrix
\begin{equation}\label{SU(2)}
  U=\left(\begin{array}{cc}b_1&-b_2^\ast\\b_2&b_1^\ast\end{array}\right)\ \ \ \
  ({\rm where}\ |b_1|^2+|b_2|^2=1)
\end{equation}
on the doublet of scalars $\phi_1$ and $\phi_2$, i.e. $\phi_a\rightarrow U_{ab}\phi_b$,
one obtains a more general solution
\begin{equation}\label{rotated-n2=0}
  \phi_1=\left(\begin{array}{cc}b_1\psi&0\\-b_2^\ast\chi&0\end{array}\right)\ ,
  \ \ \phi_2=\left(\begin{array}{cc}b_2\psi&0\\b_1^\ast\chi&0\end{array}\right)\ .
\end{equation}
The constant phases of $\psi,\chi$ in the previous solution
become part of the `moduli' $b_1,b_2$.

Having a classical moduli space in the solution, we semi-classically (or geometrically)
quantize them by computing the symplectic 2-form on the solution space, generated by
$b_1,b_2$ satisfying $|b_1|^2\!+\!|b_2|^2\!=\!1$. We ignore the last constraint for a while,
which shall be imposed later after quantization as an operator constraint. With the
canonical momenta $\pi_{\phi_1}=D_t\phi_1^\dag$, $\pi_{\phi_2}=D_t\phi_2^\dag$,
the symplectic 2-form on the space of fields is given by
\begin{equation}
  \omega=\int_{S^2}{\rm tr}\left(\delta\phi_1\wedge\delta\pi_{\phi_1}+
  \delta\phi_2\wedge\delta\pi_{\phi_2}+c.c.\right)\ ,
\end{equation}
where the exterior derivatives $\delta$ are taken in the phase space of classical fields.
Now we restrict the above 2-form $\omega$ to the solution space spanned by $b_1,b_2$.
The exterior derivatives are taken in the space $\mathbb{C}^2$
spanned by them. By using (\ref{gauss-n2=0}), one can easily show that
\begin{equation}
  \omega=ik(n_1\!-\!n_2)\delta b_a\wedge
  \delta b_a^\ast\ .
\end{equation}
This implies that
\begin{equation}
  \sqrt{\frac{1}{k(n_1\!-\!n_2)}}\ \ b_a
\end{equation}
for $a=1,2$ are annihilation operators of the two dimensional
harmonic oscillator system. Denoting by $N_a$ the corresponding occupation numbers,
the constraint on the variables $b_a$ amounts to an operator constraint
\begin{equation}\label{total-occupation}
  N_1+N_2=k(n_1\!-\!n_2)\ .
\end{equation}
Classically, the solution space subject to the constraint in (\ref{SU(2)}) is real
3 dimensional. In particular, a phase rotating $b_1,b_2$ together survives this constraint.
However, since the conjugate momentum $N_1\!+\!N_2$ of this phase is
constrained as (\ref{total-occupation}), this phase is maximally uncertain quantum
mechanically and we are effectively left with a $\mathbb{CP}^1$ phase space.
Since the oscillators $b_1^\dag$ and $b_2^\dag$ form a doublet of $SU(2)_R$,
excitation of a quantum of first and second type carries the Cartan charge
$\pm\frac{1}{2}$, respectively. Denoting by $r$ its chemical potential, the
partition function for the ground states is simply
$\chi_{k(\tilde{n}_1\!-\!\tilde{n}_2)}(r)$, exactly agreeing with the result
(\ref{U(2)-index-n2=0}) from the index.

\begin{figure}[t!]
  \begin{center}
    \includegraphics[width=6cm]{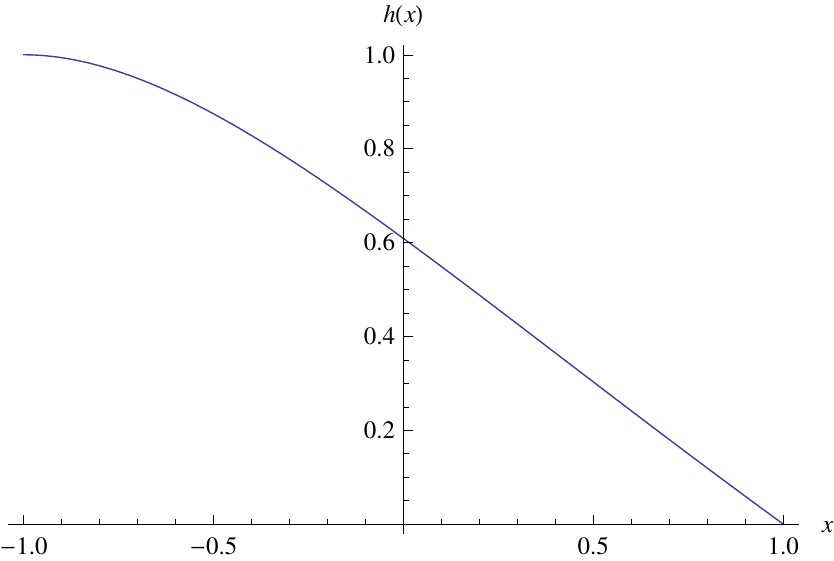}\hspace{2cm}
    \includegraphics[width=6cm]{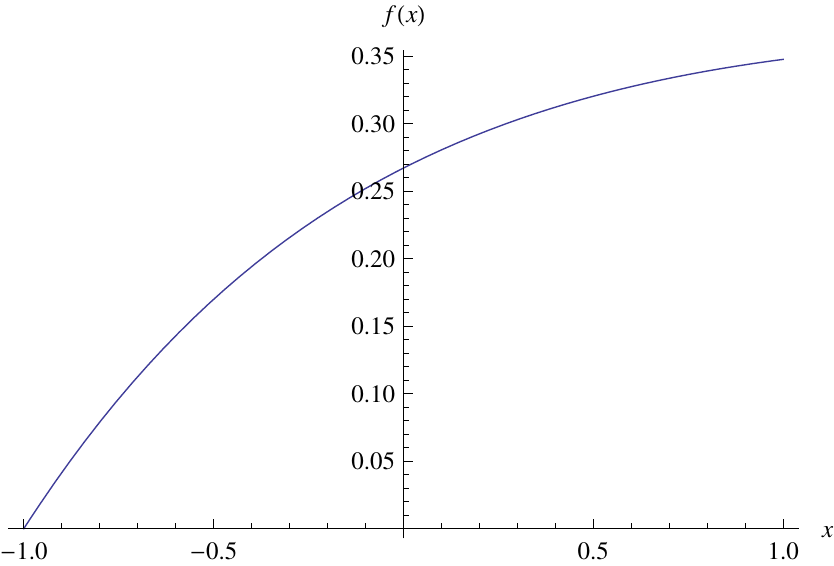}
    \includegraphics[width=6cm]{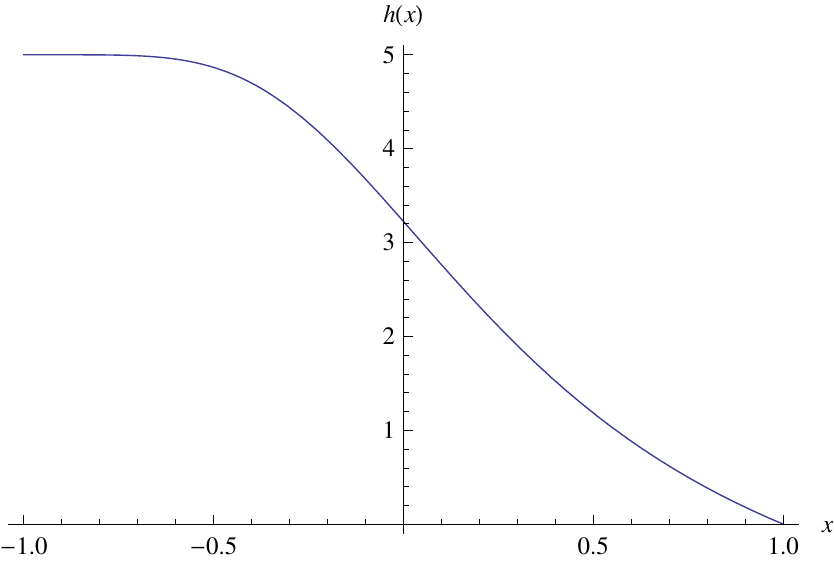}\hspace{2cm}
    \includegraphics[width=6cm]{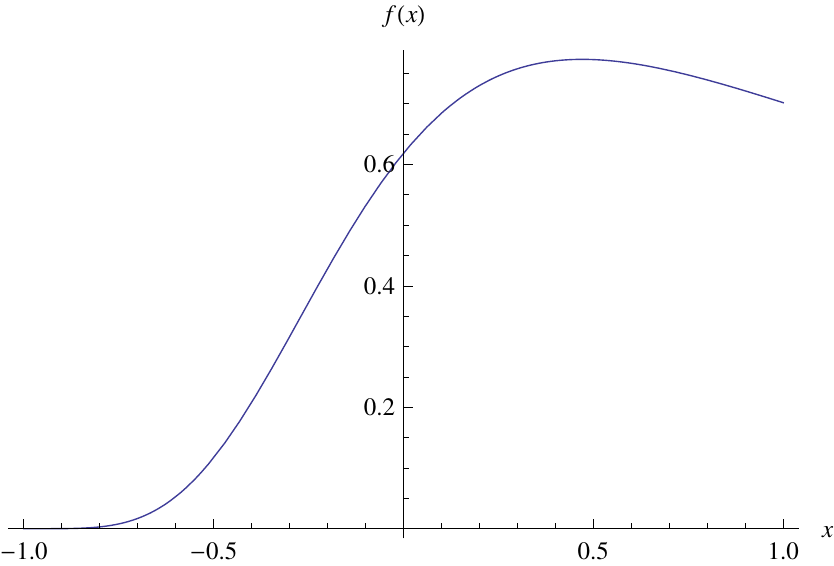}
    \includegraphics[width=6cm]{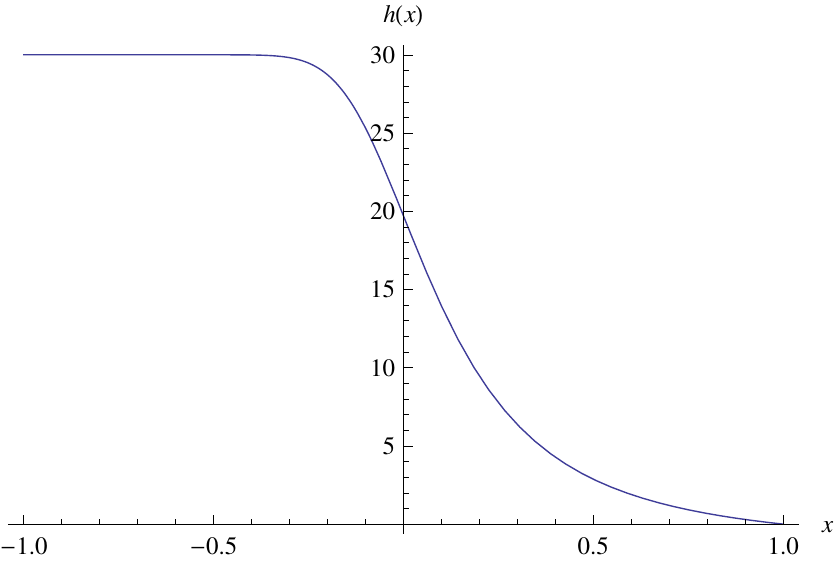}\hspace{2cm}
    \includegraphics[width=6cm]{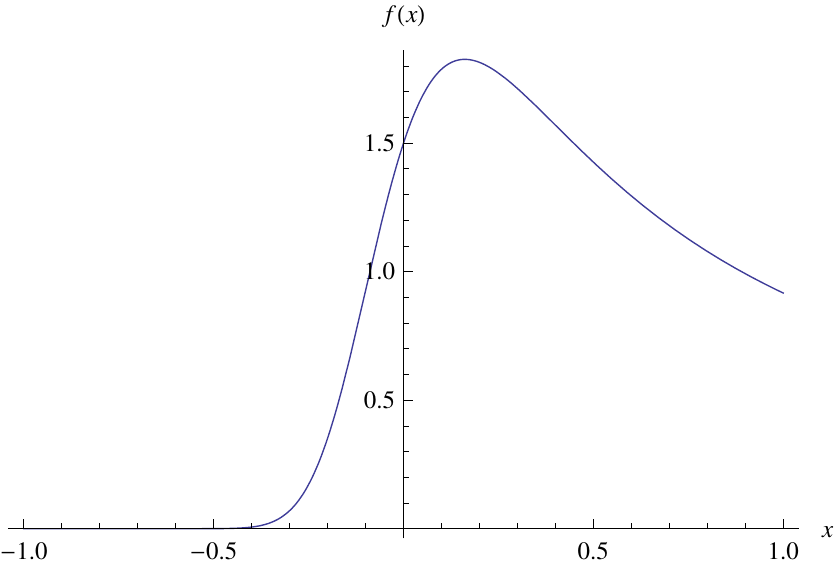}
    \includegraphics[width=6cm]{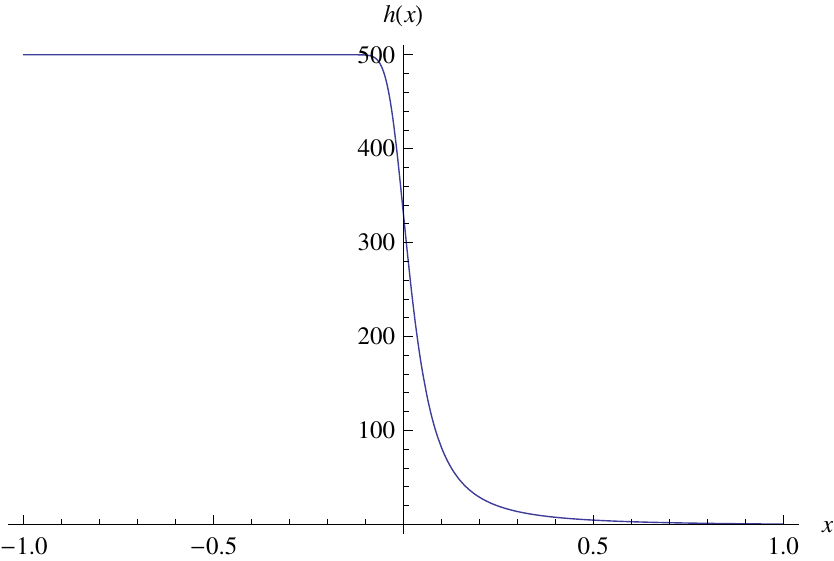}\hspace{2cm}
    \includegraphics[width=6cm]{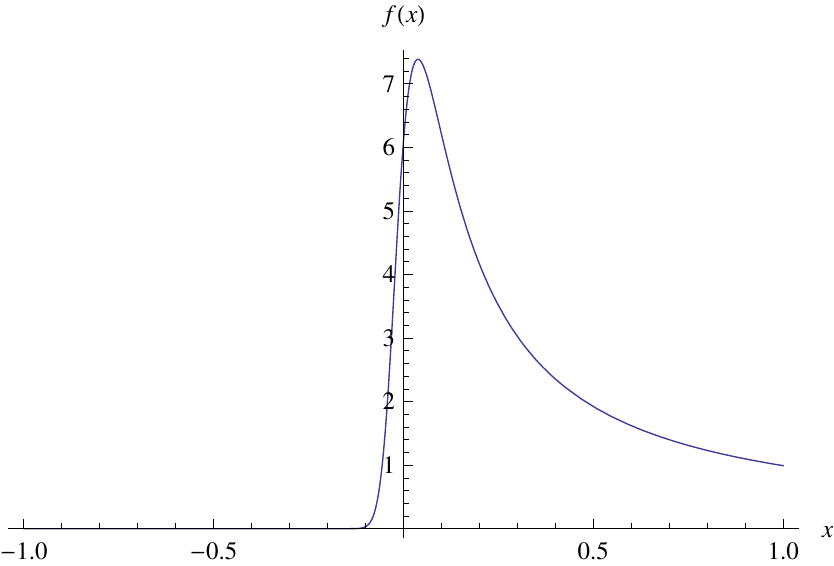}
\caption{Profiles of $h(x)$ and $f(x)$ for $H\!=\!(2n,0)$,
$\tilde{H}\!=\!(n,n)$. The fluxes are
$n\!=\!h(-1)\!\cong\!1,5,30,500$ (from top to bottem).
As the uniform flux is given by $h(x)=\frac{n}{2}(1\!-\!x)$,
one can see that magnetic flux gets more concentrated to the
equator of $S^2$ as $n$ increases.}\label{simplest}
  \end{center}
\end{figure}

A special case with $n_1\!=\!n_2$ is worth an emphasis.
In this case, the fields are naturally restricted as $g_1=g_2$,
$h_1=h_2$, $\psi=\chi$. As we know from the study above that there are no
moduli in the solution for fixed $n_1,n_2$ apart from those generated by
$SU(2)_R$, there cannot be solutions other than those obtained with this
restriction. In particular, with these restrictions,
the second $U(2)$ in the
$U(2)\times U(2)$ gauge symmetry is unbroken. From this, one can see that
the generalized solution obtained by an $SU(2)_R$ action is equivalent to
the original solution by an action of the global part of this unbroken
gauge symmetry. To see this, note that the second $U(2)$ gauge transformation
acts on the scalars from the left:
\begin{equation}
  \phi_a\rightarrow\left(\begin{array}{cc}c_1&-c^\ast_2\\c_2&c_1^\ast\end{array}\right)
  \phi_a\ \ \ ({\rm where}\ |c_1|^2\!+\!|c_2|^2\!=\!1):\
  \phi_1\rightarrow\left(\begin{array}{cc}c_1\psi&0\\c_2\psi&0\end{array}\right)\ ,\ \
  \phi_2\rightarrow\left(\begin{array}{cc}-c_2^\ast\chi&0\\c_1^\ast\chi&0\end{array}
  \right)\ .
\end{equation}
From $\psi\!=\!\chi$ for $n_1\!=\!n_2$, this takes the same form as the action of
$SU(2)_R$ global symmetry (\ref{rotated-n2=0}), proving our claim. Therefore, the unbroken
gauge symmetry eliminates the `would-be moduli' as a gauge orbit. The
absence of classical moduli, and thus the absence of the phase
space, implies that there is no ground state degeneracy. This is
consistent with the formula from the index for
$\tilde{n}_1\!=\!\tilde{n}_2$, as
$\chi_{k(\tilde{n}_1\!-\!\tilde{n}_2)}(r)$ reduces to $\chi_0(r)=1$.

In this case, namely with $H=(2n,0)$, $\tilde{H}=(n,n)$, the
equations for $g\equiv g_1=g_2$, $h\equiv h_1=h_2$, $f\equiv
f_1=f_2$ reduce to
\begin{equation}
  h^\prime=2gf\ ,\ \ (1-x^2)g^\prime=2hf\ ,\ \ xf=-\left(g+h+\frac{1}{2}\right)\ .
\end{equation}
With $n>0$, one again finds that the boundary conditions should be
$h(1)=0$ and $f(-1)=0$, for the solution to be regular. From (\ref{x=1-BC}),
the free boundary parameter determining the flux $n=h(-1)$ is $g(1)$.
Since we have only one free parameter to control, we can systematically
search for the allowed range of $g(1)$ and see how the flux $n$
depends on it. Decreasing $g(1)$ in the range
$-\frac{1}{2}\!>\!g(1)\!>\!-\frac{3}{2}$, we find solutions with
increasing magnetic charge in $0\!<\!n\!<\!\infty$. In particular,
we can find solutions with arbitrarily large $n$ by taking $g(1)$ to
be close to $-\frac{3}{2}$. Fig.\ref{simplest} shows the profiles of
the magnetic potential $h(x)$ and the scalar-squared $f(x)$ for
$g(1)=-0.8477,-1.20163,-1.41641,-1.491257$. The corresponding fluxes
are $n\!=\!1.00003$, $4.9999$, $30.0045$, $499.997$, tuned to be close
to $n\cong 1,5,30,500$, respectively.

\subsection{$U(2)\times U(2)$ monopoles with general magnetic charges}

Now we turn to construct and study solutions with general magnetic
charges $H=(n_1,n_2)$, $\tilde{H}=(\tilde{n}_1,\tilde{n}_2)$. The
analysis for this general case will be much more
complicated than the special case with $n_2\!=\!0$ in the previous
subsection. In particular, it will turn out that the solution we
construct in this section will only provide a subset of the most
general solution.

Following the previous subsection,
we take an ansatz in which gauge fields are diagonal,
\begin{equation}
  A_t=\left(\begin{array}{cc}g_1&0\\0&g_2\end{array}\right),\
  \tilde{A}_t=\left(\begin{array}{cc}\tilde{g}_1&0\\0&\tilde{g}_2\end{array}\right),\
  A_\varphi=\left(\begin{array}{cc}h_1&0\\0&h_2\end{array}\right),\ \tilde{A}_\varphi
  =\left(\begin{array}{cc}\tilde{h}_1&0\\0&\tilde{h}_2\end{array}\right)\ ,
\end{equation}
with $A_\theta\!=\!\tilde{A}_\theta\!=\!0$. The components are again
taken to be functions of $x=\cos\theta$ only. As we shall see
shortly from the semi-classical quantization of the moduli, this
ansatz seems to be too restrictive to reproduce the full
degeneracy that we studied from the index. Presumably, we should
relax this ansatz by allowing nonzero off-diagonal elements. We
leave this generalization as a future work, and concentrate on the study
of the solutions within this ansatz. The functions are restricted
as
\begin{equation}
  h_1(-1)\!-\!h_1(1)\!=\!n_1,\ h_2(-1)\!-\!h_2(1)\!=\!n_2,\
  \tilde{h}_1(-1)\!-\!\tilde{h}_1(1)\!=\!\tilde{n}_1,\
  \tilde{h}_2(-1)\!-\!\tilde{h}_2(1)\!=\!\tilde{n}_2
\end{equation}
due to the flux condition. As in the previous section, we shall
consider the case in which all fluxes are positive. Up to a global
$SU(2)_R$ symmetry transformation which will generate a moduli, we find
that the scalar fields should be taken to be
\begin{equation}
  \phi_1=\left(\begin{array}{cc}\psi_1e^{im_1\varphi-i\omega_1t}&0\\0&
  \psi_2e^{im_2\varphi-i\omega_2t}\end{array}\right)
  \ ,\ \ \phi_2=\left(\begin{array}{cc}0&\chi_2e^{im_4\varphi-i\omega_4t}\\
  \chi_1e^{im_3\varphi-i\omega_3t}&0\end{array}\right)
\end{equation}
where functions again depend on $x\!=\!\cos\theta$ only. The nonzero
components of the scalars are carefully chosen so that the ansatz
automatically satisfies the off-diagonal parts of the Gauss' law with
the above diagonal gauge fields.

We comment
on the possibility of using the local $U(1)^4$ gauge transformations
in $U(2)\times U(2)$ to eliminate the $\varphi,t$ dependent phases
in the scalars. The diagonal $U(1)$ in
$U(1)^4\subset U(2)\times U(2)$ decouples to all matter fields.
Therefore, only three of the four local $U(1)^4$ can be used to eliminate
three combinations of $m_{1,2,3,4}$, and also three combinations of
$\omega_{1,2,3,4}$. In fact, it is easy to show that the following
two combinations
\begin{equation}
  m_1\!+\!m_2\!-\!m_3\!-\!m_4\ ,\ \ \omega_1\!+\!\omega_2\!-\!\omega_3\!-\!\omega_4
\end{equation}
cannot be changed by such gauge transformations. For instance, we
can set three of the four $\varphi,t$ dependent phases in the above
scalars to be $1$. Below, at suitable stage, we will set
$m_1\!=\!m_2\!=\!m_3\!=\!0$ and
$\omega_1\!=\!\omega_2\!=\!\omega_3\!=\!0$ while keeping nonzero
$m\!\equiv\!m_4$ and $\omega\!\equiv\!\omega_4$ only.

The supersymmetry condition is given by
\begin{eqnarray}
  x(1-x^2)(\psi_{1,2}^\ast)^\prime&=&\psi_{1,2}^\ast\left[(1-x^2)\left(
  \omega_{1,2}+\tilde{g}_{1,2}-g_{1,2}-\frac{1}{2}\right)-m_{1,2}+\tilde{h}_{1,2}-h_{1,2}
  \right]\nonumber\\
  x(1-x^2)(\chi_{1,2}^\ast)^\prime&=&\chi_{1,2}^\ast\left[(1-x^2)\left(
  \omega_{3,4}+\tilde{g}_{2,1}-g_{1,2}-\frac{1}{2}\right)-m_{3,4}+\tilde{h}_{2,1}-h_{1,2}
  \right]\nonumber\\
  (1-x^2)(\psi_{1,2}^\ast)^\prime&=&\psi_{1,2}^\ast
  \left[-x\left(\omega_{1,2}+\tilde{g}_{1,2}-g_{1,2}-\frac{1}{2}\right)\pm\frac{2\pi}{k}
  \left(|\chi_1|^2-|\chi_2|^2\right)\right]\nonumber\\
  (1-x^2)(\chi_{1,2}^\ast)^\prime&=&\chi_{1,2}^\ast
  \left[-x\left(\omega_{3,4}+\tilde{g}_{2,1}-g_{1,2}-\frac{1}{2}\right)\pm\frac{2\pi}{k}
  \left(|\psi_1|^2-|\psi_2|^2\right)\right]\ .
\end{eqnarray}
First of all, from these equations one can easily check that the
phases of $\psi_{1,2}$ and $\chi_{1,2}$ are all independent of
$\theta$. Also, combining the first and third equations, and also
the second and fourth equations appropriately, one obtains the
following algebraic conditions:
\begin{eqnarray}\label{algebraic-general}
  \hspace*{-0.7cm}\frac{2\pi}{k}x\left(|\psi_1|^2\!-\!|\psi_2|^2\right)\!&\!=\!&
  \!-\!\left(\!\omega_4\!-\!m_4\!+\!\tilde{g}_1\!+\!\tilde{h}_1\!-\!g_2\!-\!h_2\!-\!\frac{1}{2}
  \!\right)\!=\!
  \omega_3\!-\!m_3\!+\!\tilde{g}_2\!+\!\tilde{h}_2\!-\!g_1\!-\!h_1\!-\!\frac{1}{2}\nonumber\\
  \hspace*{-0.7cm}\frac{2\pi}{k}x\left(|\chi_1|^2\!-\!|\chi_2|^2\right)\!&\!=\!&\!
  -\!\left(\!\omega_2\!-\!m_2\!+\!\tilde{g}_2\!+\!\tilde{h}_2\!-\!g_2\!-\!h_2\!-\!
  \frac{1}{2}\!\right)\!=\!
  \omega_1\!-\!m_1\!+\!\tilde{g}_1\!+\!\tilde{h}_1\!-\!g_1\!-\!h_1\!-\!\frac{1}{2}\ .
\end{eqnarray}
In particular, the second equality on both lines demand
\begin{equation}\label{off-diagonal-U(1)}
  (g_1+g_2-\tilde{g}_1-\tilde{g}_2)+(h_1+h_2-\tilde{h}_1-\tilde{h}_2)=
  \omega_1+\omega_2-m_1-m_2-1
\end{equation}
and
\begin{equation}\label{relation-frequency}
  \omega_1+\omega_2-m_1-m_2=\omega_3+\omega_4-m_3-m_4\ .
\end{equation}
The terms in the first and second parentheses on the left hand side of
(\ref{off-diagonal-U(1)}) are ${\rm tr}(A_t\!-\!\tilde{A}_t)$ and
${\rm tr}(A_\varphi\!-\!\tilde{A}_\varphi)$, which we know should be
separately set to be constants from decoupling of an overall $U(1)$.
This should also be manifest in the Gauss' law we study below.
Therefore, in foresight, we set
\begin{equation}\label{decoupled}
  h_1+h_2-\tilde{h}_1-\tilde{h}_2=c\ ,\ \
  g_1+g_2-\tilde{g}_1-\tilde{g}_2=-1-c+(\omega_1+\omega_2-m_1-m_2)
\end{equation}
with a constant $c$. As we shall see below, the value of $c$ is not
determined at this state, but should be appropriately chosen to
yield the correct ground state energy.\footnote{ We emphasize that,
the choice of our ansatz does not yet force us to consider the lowest
energy configurations with given magnetic charges. For instance, in
the index studied in \cite{Kim:2009wb}, the ground states come from
exciting lowest monopole spherical harmonics to saturate Gauss' law,
while higher spherical harmonics with larger angular momenta give
gauge invariant states with higher energies.}

To summarize up to now, the algebraic and differential conditions
from supersymmetry are
\begin{eqnarray}\label{susy}
  x\left(f_1-f_2\right)&=&-\left(\omega_4-m_4+
  \tilde{g}_1+\tilde{h}_1-g_2-h_2-\frac{1}{2}\right)\nonumber\\
  x\left(f_3-f_4\right)&=&-\left(\omega_2-m_2+
  \tilde{g}_2+\tilde{h}_2-g_2-h_2-\frac{1}{2}\right)\nonumber\\
  (1-x^2)f_{1,2}^\prime&=&2f_{1,2}\left[-x\left(\omega_{1,2}+\tilde{g}_{1,2}-g_{1,2}
  -\frac{1}{2}\right)\pm\left(f_3-f_4\right)
  \right]\nonumber\\
  (1-x^2)f_{3,4}^\prime&=&2f_{3,4}\left[-x\left(\omega_{3,4}+\tilde{g}_{2,1}-
  g_{1,2}-\frac{1}{2}\right)\pm\left(f_1-f_2\right)\right]\ ,
\end{eqnarray}
where we defined $f_1\equiv\frac{2\pi}{k}|\psi_1|^2$,
$f_2\equiv\frac{2\pi}{k}|\psi_2|^2$, $f_3=\frac{2\pi}{k}|\chi_1|^2$,
$f_4=\frac{2\pi}{k}|\chi_2|^2$, and the parameters are subject to
the condition (\ref{relation-frequency}). In addition, phases of
$\psi_{1,2}$ and $\chi_{1,2}$ are all constants.

The Gauss' law conditions are given by the following eight equations:
\begin{eqnarray}\label{gauss}
  h_{1}^\prime&=&2f_1(g_1-\tilde{g}_1-\omega_1)+2f_3(g_1-\tilde{g}_2-\omega_3)\nonumber\\
  h_{2}^\prime&=&2f_2(g_2-\tilde{g}_2-\omega_2)+2f_4(g_2-\tilde{g}_1-\omega_4)\nonumber\\
  \tilde{h}_{1}^\prime&=&2f_1(g_1-\tilde{g}_1-\omega_1)+2f_4(g_2-\tilde{g}_1-\omega_4)\nonumber\\
  \tilde{h}_{2}^\prime&=&2f_2(g_2-\tilde{g}_2-\omega_2)+2f_3(g_1-\tilde{g}_2-\omega_3)\nonumber\\
  (1-x^2)g_{1}^\prime&=&2f_1(h_1-\tilde{h}_1+m_1)+2f_3(h_1-\tilde{h}_2+m_3)\nonumber\\
  (1-x^2)g_{2}^\prime&=&2f_2(h_2-\tilde{h}_2+m_2)+2f_4(h_2-\tilde{h}_1+m_4)\nonumber\\
  (1-x^2)\tilde{g}_{1}^\prime&=&2f_1(h_1-\tilde{h}_1+m_1)+2f_4(h_2-\tilde{h}_1+m_4)
  \nonumber\\
  (1-x^2)\tilde{g}_{2}^\prime&=&2f_2(h_2-\tilde{h}_2+m_2)
  +2f_3(h_1-\tilde{h}_2+m_3)\ .
\end{eqnarray}
Among them, a combination of the first four equations says
$(h_1\!+\!h_2\!-\!\tilde{h}_1\!-\!\tilde{h}_2)^\prime=0$, while a
combination of the last four equations says
$(g_1\!+\!g_2\!-\!\tilde{g}_1\!-\!\tilde{g}_2)^\prime=0$. As
asserted previously, these amount to the constancy of the gauge
field ${\rm tr}(A_\mu\!-\!\tilde{A}_\mu)$, which are solved
as (\ref{decoupled}). We thus have six independent equations from
the Gauss' law.

Let us consider the relation between the four differential
conditions in (\ref{susy}) and six independent equations in
(\ref{gauss}). Firstly, by adding the supersymmetry conditions
containing $f_1^\prime$ and $f_3^\prime$, one obtains
\begin{equation}
  \left(xf_1+xf_3\right)^\prime=-\left(g_1+h_1\right)^\prime
\end{equation}
after using appropriate Gauss' law equations.
Similarly, considering various different combinations
and using Gauss' law, one obtains
\begin{eqnarray}\label{solve-susy}
  &&xf_1+xf_3=-(g_1+h_1+\gamma)\ ,\ \ xf_2+xf_4=-(g_2+h_2+\gamma+\omega_3-\omega_2-m_3+m_2)\ ,
  \nonumber\\
  &&xf_1+xf_4=-\left(\tilde{g}_1+\tilde{h}_1+\gamma+\omega_1-m_1-\frac{1}{2}\right)\ .
\end{eqnarray}
The three integration constants are restricted as above from the
algebraic conditions. From the above equations, one can also obtain
$xf_2+xf_3=-(\tilde{g}_2\!+\!\tilde{h}_2\!+\!\gamma\!+\!\omega_3\!-\!m_3\!-\!\frac{1}{2})$.
These conditions imply that three of the four differential
supersymmetry conditions are guaranteed by the Gauss' law condition.

One the other hand, one can take two different combinations of
(\ref{susy}) to obtain
\begin{eqnarray}
  (1-x^2)\left(\log f_1f_2\right)^\prime&=&-2x\left(\omega_1\!+\!\omega_2\!+\!
  \tilde{g}_1\!+\!\tilde{g}_2\!-\!g_1\!-\!g_2\!-\!1\right)=-2(c\!+\!m_1\!+\!m_2)x\ ,
  \nonumber\\
  (1-x^2)\left(\log f_3f_4\right)^\prime&=&-2(c\!+\!m_3\!+\!m_4)x\ ,
\end{eqnarray}
whose solutions are
\begin{equation}\label{scalar-solution}
  f_1f_2=A(1-x^2)^{c\!+\!m_1\!+\!m_2}\ ,\ \ f_3f_4=B(1-x^2)^{c\!+\!m_3\!+\!m_4}
\end{equation}
with constant $A,B$. From these results, we expect
$c\!+m_1\!+\!m_2$ and $c\!+\!m_3\!+\!m_4$ to be positive.
Note that $\psi_1\psi_2$ and $\chi_1\chi_2$ are combinations
neutral under the magnetic field.

As we emphasized earlier, $c$ and one gauge-invariant combination of
$m_{1,2,3,4}$ are not fixed by any reason yet. However, we can
constrain them by demanding that the solution takes lowest energy in
the sector with given magnetic charges. In the previous subsection,
with $n_2\!=\!0$, our ansatz always provided solutions with
lowest energy. In the general case, we have more parameters
$c,m_{1,2,3,4}$ in the solution, which will turn out to allow
solutions with either lowest or excited energy. In the
considerations below, we shall mostly assume the values expected for
ground states and show that the solutions satisfy all the desired
properties.\footnote{For other choices of
$c,m_{1,2,3,4}$, we also found a class of excited solutions.}
Firstly, from (\ref{scalar-solution}), we expect that the conditions
for the solutions to carry lowest angular momenta are
\begin{equation}\label{parameter-ground}
  c\!+\!m_1\!+\!m_2=\tilde{n}_2\!-\!n_2\ ,\ \ c\!+\!m_3\!+\!m_4=
  \tilde{n}_1\!-\!n_2\ .
\end{equation}
To argue this, recall from the previous paragraph that $\psi_1\psi_2$ and
$\chi_1\chi_2$ are neutral under the magnetic field. They can thus be understood with
our intuition on ordinary spherical harmonics. In fact, the two right hand sides of
the solutions (\ref{scalar-solution}) are squares of the BPS spherical harmonics
\begin{equation}
 Y_{jj}=\left(\sin\theta\right)^j
\end{equation}
with $j\!=\!c\!+m_1\!+\!m_2$ and $j\!=\!c\!+m_3\!+\!m_4$.
As the fields $\psi_1,\psi_2$ feel the magnetic charges
$\tilde{n}_1\!-\!n_1$ and $\tilde{n}_2\!-\!n_2\!=\!-(n_1\!-\!\tilde{n}_1)$
while $\chi_1,\chi_2$ feel $\tilde{n}_2\!-\!n_1$ and $\tilde{n}_1\!-\!n_2\!=\!-
(\tilde{n}_2\!-\!n_1)$, the products $\psi_1\psi_2$ and $\chi_1\chi_2$ are expected
to carry minimal angular momenta $\tilde{n}_2\!-\!n_2$ and $\tilde{n}_1\!-\!n_2$,
respectively, leading to (\ref{parameter-ground}). Since angular momentum contributes
to BPS energy, having minimal angular momenta is part of the requirement for the
ground state solutions. Furthermore, when $n_1\!>\!\tilde{n}_2\!>\!\tilde{n}_1\!>\!n_2$,
we have explained in section 2 that the $12$
matrix element of the anti-bifundamental scalars should be zero to have
lowest energy. In our ansatz, this amounts to taking
\begin{equation}\label{zero-scalar}
  f_4=0\ .
\end{equation}
The two conditions (\ref{parameter-ground}) and (\ref{zero-scalar})
are the condition that we impose by hand to obtain the ground state
solutions for generic fluxes. From the conserved charges that we calculate below,
the relevance of these conditions for the ground states will be manifest.

Let us calculate the Noether charges of the solution.
The $U(1)_R$ charge $q$ entering in the BPS energy condition is given by
\begin{equation}\label{R-charge-general}
  q=2\pi\int_{-1}^1 dx{\rm tr}\left(\frac{i}{2}D_t\phi_a\phi_a^\dag+c.c.\right)
  =-\frac{k}{2}\int_{-1}^1dx\left(\tilde{h}_1^\prime+\tilde{h}_2^\prime\right)
  =\frac{k}{2}\left(\tilde{n}_1+\tilde{n}_2\right)\ ,
\end{equation}
which is compatible with the expectation from the index. On the other hand,
after a bit lengthy manipulation using (\ref{gauss}) and $h_1\!+\!h_2\!-\!\tilde{h}_1\!-\!
\tilde{h}_2\!=\!c$, the Noether angular momentum
is given by
\begin{equation}\label{angular-general}
  \hspace*{-0.8cm}j\!=\!-k\!\int_{-1}^1\!\!dx\!\left[\left((-m_1\!+\!\tilde{h}_1\!-\!h_1)
  (-m_3\!+\!\tilde{h}_2\!-\!h_1)+(m_1\!+\!m_2\!+\!c)h_2\!\right)^\prime\!\!
  +\!2(m_1\!+\!m_2\!-\!m_3\!-\!m_4)(\omega_4\!+\!\tilde{g}_1\!-\!g_2)f_4\right]\ .
\end{equation}
After a much lengthier calculation, one can also show explicitly
that the Noether energy satisfies the BPS relation
$\epsilon\!=\!q\!+\!j$. From (\ref{angular-general}), we find that
the angular momentum and the BPS energy depend on the details of the
solution, contrary to the R-charge (\ref{R-charge-general}) which is
determined by the monopole charges only. The reason for this is that
our ansatz actually can cover monopoles with excited energies above the
ground states. To study the conserved charges for the ground states,
we apply the conditions (\ref{parameter-ground}) and
(\ref{zero-scalar}) that we expect for the classical solutions with
lowest energy. From $f_4\!=\!0$ and
$m_1\!+\!m_2\!+\!c\!=\!\tilde{n}_2\!-\!n_2$ for the ground states,
$j$ in (\ref{angular-general}) is given in terms of the boundary
values of the functions at $x\!=\!\pm 1$. Or when $\tilde{n}_1\!=\!\tilde{n}_2$,
for which $f_4\!=\!0$ need not be imposed, (\ref{parameter-ground}) says
$m_1\!+\!m_2\!-\!m_3\!-\!m_4\!=\!0$, eliminating the second term in
(\ref{angular-general}). As we shall argue later
when we discuss the ground state solutions, the boundary conditions
for regular solutions at the north pole $x\!=\!1$ should be
$\tilde{h}_1(1)\!-\!h_1(1)\!=\!m_1$,
$\tilde{h}_2(1)\!-\!h_1(1)\!=\!m_3$ for
$n_1\!>\!\tilde{n}_1\!\geq\!\tilde{n}_2\!>\!n_2\!>\!0$.
Imposing these conditions, the angular momentum is given by
\begin{eqnarray}
  j&=&-k\int_{-1}^1dx\left[(-m_1\!+\!\tilde{h}_1\!-\!h_1)(-m_3\!+\!
  \tilde{h}_2\!-\!h_1)+(\tilde{n}_2-n_2)h_2\right]^\prime\nonumber\\
  &=&k\left[(-m_1\!+\!\tilde{h}_1(-1)\!-\!h_1(-1))
  (-m_3\!+\!\tilde{h}_2(-1)\!-\!h_1(-1))+
  (\tilde{n}_2\!-\!n_2)\left(h_2(-1)\!-\!h_2(1)\right)\right]\nonumber\\
  &=&k\left[\frac{}{}(n_1\!-\!\tilde{n}_1)(n_1\!-\!\tilde{n}_2)+
  (\tilde{n}_2\!-\!n_2)n_2\right]=k\tilde{n}_1(\tilde{n}_2-n_2)\ .
\end{eqnarray}
The last expression is exactly the angular momentum of ground states
(\ref{charge-index}) that we obtained from the index.

Now we explain how to construct numerical solutions. We shall first
consider the case with $n_1\!>\!\tilde{n}_1\!>\!\tilde{n}_2\!>\!n_2$, and
then the case with $n_1\!>\!\tilde{n}_1\!=\!\tilde{n}_2\!>\!n_2$. These two cases,
together with the case $n_1\!=\!\tilde{n}_1\!\geq\!\tilde{n}_2\!=\!n_2$
discussed in \cite{Kim:2009ia}, essentially exhaust the most general flux
in $U(2)\times U(2)$. By setting $m_1\!=\!m_2\!=\!m_3\!=\!0$ and
$\omega_1\!=\!\omega_2\!=\!\omega_3\!=\!0$ at this stage, we obtain
$c\!=\!\tilde{n}_2\!-\!n_2$ and
$m\!\equiv\!m_4\!=\!\tilde{n}_1\!-\!\tilde{n}_2\!=\!\omega_4\!\equiv\!\omega$.

We start from the case $n_1\!>\!\tilde{n}_1\!>\!\tilde{n}_2\!>\!n_2$
with (\ref{parameter-ground}) and (\ref{zero-scalar}) satisfied. Since $f_4\!=\!0$,
$f_1,f_2,f_3$ are determined in terms of the functions from gauge fields via
(\ref{solve-susy}). We can set $\gamma\!=\!0$ by using, say $t$ dependent gauge
transformation in overall $U(1)$ which shifts $g_1,g_2,\tilde{g}_1,\tilde{g}_2$
altogether. The expressions are given by
\begin{equation}
  f_1=-\frac{\tilde{g}_1+\tilde{h}_1-\frac{1}{2}}{x}\ ,\ \ f_2=-\frac{g_2+h_2}{x}
  \ ,\ \ f_3=\frac{\tilde{g}_1+\tilde{h}_1-g_1-h_1-\frac{1}{2}}{x}\ .
\end{equation}
We can take six independent functions $h_1,h_2,\tilde{h}_1$, $g_1,g_2,\tilde{g}_1$
to be determined numerically, where $\tilde{h}_2$, $\tilde{g}_2$ can be written
in terms of these six functions using (\ref{decoupled}). The six equations determining
the independent functions can be taken to be the first, second, third, fifth, sixth and
seventh equations of (\ref{gauss}). We should first specify
the correct boundary conditions at $x\!=\!\pm 1$ to have regular solutions,
because the equations containing $(1\!-\!x^2)$ on the left hand sides
of (\ref{gauss}) should have the corresponding right hand sides to be zero
at $x\!=\!\pm 1$, like the case of previous subsection. Now
recall that the modes $\psi_1$, $\chi_1$ feel negative magnetic fluxes
$\tilde{n}_1\!-\!n_1$, $\tilde{n}_2\!-\!n_1$, respectively, while $\psi_2$, $\chi_2$
feel positive fluxes $\tilde{n}_2\!-\!n_2$, $\tilde{n}_1\!-\!n_2$, respectively.
Therefore, similar to the arguments in the previous subsection, $f_1$, $f_3$ should be
nonzero at $x\!=\!1$ while $f_2$ should be nonzero at $x\!=\!-1$.
From the right hand sides of fifth, sixth and seventh equations
of (\ref{gauss}), one obtains
\begin{equation}
  g_2(1)=-h_1(1)-c\ ,\ \
  h_2(1)=h_1(1)+c\ ,\ \ \tilde{h}_1(1)=h_1(1)
\end{equation}
with $h_1(1)$, $g_1(1)$, $\tilde{g}_1(1)$ unconstrained at $x\!=\!1$, and
\begin{equation}
 \tilde{h}_1(-1)=h_1(-1)-c\ ,\ \ g_1(-1)=-h_1(-1)\ ,\ \
 \tilde{g}_1(-1)=\frac{1}{2}+c-h_1(-1)
\end{equation}
with $h_1(-1),h_2(-1),g_2(-1)$ (yet) unconstrained at $x\!=\!-1$.

Firstly, we consider the regular asymptotic solution at $x\!=\!1$.
We expand the six functions with $y\equiv 1-x^2$ near $x\!=\!1$ and solve
the independent equations in (\ref{gauss}). After some calculation,
one obtains
\begin{eqnarray}\label{expansion-x=1}
  h_1(x)&=&a_1+y\ \left(-\frac{1}{2}-\frac{3}{2}a_2+2a_3+2a_2a_3
  -2a_3^2+\frac{1}{2}a_1+2a_1a_2-2a_1a_3\right)+\cdots\nonumber\\
  h_2(x)&=&c+a_1+\frac{a_4(-1-c-a_2+a_3)}{1+c}\ y^{1+c}+\cdots\nonumber\\
  \tilde{h}_1(x)&=&a_1+y\ \left(-\frac{1}{2}a_2+\frac{1}{2}a_3+a_2a_3-a_3^2
  +a_1a_2-a_1a_3\right)+\cdots\nonumber\\
  g_1(x)&=&a_2+y\ \left(\frac{1+a_2-2a_2^2}{4}-a_3+a_2^2a_3+a_3^2-a_2a_3^2
  -\frac{3}{4}a_1-a_1a_2+a_1a_2^2+2a_1a_3-a_1a_3^2\right.\nonumber\\
  &&\hspace{2cm}\left.+\frac{a_1^2}{2}+a_1^2a_2-a_1^2a_3\right)+\cdots\\
  g_2(x)&=&-c-a_1+a_4y^c+\cdots\nonumber\\
  \tilde{g}_1(x)&=&a_3+y\ \left(\frac{1}{4}+\frac{a_2}{2}-\frac{5a_3}{4}-\frac{3a_2a_3}{2}
  +2a_3^2+a_2a_3^2-a_3^3-\frac{3a_1}{4}-\frac{3a_1a_2}{2}\right.\nonumber\\
  &&\hspace{2cm}\left.+\frac{5a_1a_3}{2}+2a_1a_2a_3-2a_1a_3^2+\frac{a_1^2}{2}+a_1^2a_2
  -a_1^2a_3\right)+\cdots\nonumber
\end{eqnarray}
with $4$ independent coefficients $a_1=h_1(1)$, $a_2=g_1(1)$, $a_3=\tilde{g}_1(1)$,
$a_4$.

We also consider regular asymptotic solution at $x\!=\!-1$. There turn out to be
many possible asymptotic expansions at $x\!=\!-1$, due to subtle factorizations.
Depending on the values of the fluxes $H,\tilde{H}$, different expansion would be
relevant. As an illustration, let us present an expansion which would be relevant
for one of our numerical solutions below. There are four independent parameters
in this expansion, $b_1=h_1(-1)$, $b_2=g_2(-1)$, $b_3$ and $b_4$, and the solution
near $x\!=\!-1$ is
\begin{eqnarray}\label{expansion-x=-1}
  h_1(x)&=&b_1-b_3\frac{\frac{1}{2}+c}{1+c}\ y^{c+1}+\mathcal{O}(y^{c+2})\nonumber\\
  h_2(x)&=&b_1-1-c-\frac{b_1+b_2-1-c}{2}\ y-\frac{b_1+b_2-1-c}{8}\ y^2+\mathcal{O}(y^3)\nonumber\\
  \tilde{h}_1(x)&=&b_1-c-b_3\frac{\frac{1}{2}+c}{1+c}\ y^{c+1}+\mathcal{O}(y^{c+2})\nonumber\\
  g_1(x)&=&-b_1+b_3y^c+b_4y^{c+1}+\mathcal{O}(y^{c+2})\nonumber\\
  g_2(x)&=&b_2+0\cdot y^{c+1}+\mathcal{O}(y^{c+2})\nonumber\\
  \tilde{g}_1(x)&=&\frac{1}{2}+c-b_1+b_3y^c+b_3\ \frac{c}{2+2c}y^{c+1}+\mathcal{O}(y^{c+2})
\end{eqnarray}
where all omitted terms are determined by $b_1,b_2,b_3,b_4$.
In our gauge, we have $h_2(1)\!-\!h_1(1)\!=\!c$ from (\ref{expansion-x=1}). The expansion
at $x\!=\!-1$ that we show in (\ref{expansion-x=-1}) have $h_2(-1)\!-\!h_1(-1)\!=\!-1\!-\!c$,
which implies
\begin{equation}\label{flux-constraint}
  n_1\!-\!n_2=h_1(-1)-h_1(1)-h_2(-1)+h_2(1)=2c+1\ .
\end{equation}
Below, we will construct numerical solutions with $H\!=\!(4,1)$, $\tilde{H}\!=\!(3,2)$.
As $c\!=\!\tilde{n}_2\!-\!n_2$ should be taken to be $1$, (\ref{flux-constraint}) is
satisfied with these fluxes. In general, $n_1\!-\!n_2$ is different from $2c\!+\!1=2\tilde{n}_2\!-\!2n_2\!+\!1$. We find that there are
so many branches of possible expansions at $x\!=\!-1$ that one can choose appropriate
$h_2(-1)$ to fit the flux that one desires to have. As this study is very cumbersome,
we have not carried out the full analysis in general. We experienced that, in all
branches we studied, there always exist four independent parameters at $x\!=\!-1$
(like our $b_1,b_2,b_3,b_4$ above).

In total, we have eight parameters appearing in the regular asymptotic solutions
at both ends $x\!=\!\pm 1$. A practical way of viewing the parameters in the solutions,
which will also be useful for understanding numerical analysis below,
can be summarized as follows. Firstly, one picks a set of values for $a_1,a_2,a_3,a_4$
satisfying regularity condition at $x\!=\!1$. As we solve the six differential equations
to $x\!=\!-1$, a generic choice of the 4 parameters at $x\!=\!1$ will violate
the regularity condition at $x\!=\!-1$. A 2-parameter fine-tuning at $x\!=\!1$
would have the regularity condition at $x\!=\!-1$ satisfied, as we have four
$b_1,b_2,b_3,b_4$ there.\footnote{We shall see below from a $U(1)$ symmetry that
the fine-tuning is actually 1-dimensional.} Then, the remaining two parameters (after tuning)
among $a_1,a_2,a_3,a_4$ are left, which also determines $b_{1,2,3,4}$. The last two
parameters should determine two of the three independent fluxes
(subject to $n_1\!+\!n_2\!=\!\tilde{n}_1\!+\!\tilde{n}_2$). Another
independent flux is not encoded in the boundary conditions but
is chosen by fixing $c$ to be $\tilde{n}_2\!-\!n_2\!=\!c$. Recall that
this relation was made to have lowest angular momentum states: technically,
the last condition was imposed by demanding nonzero values of $f_1,f_3$
and $f_2$ at $x\!=\!1$ and $x\!=\!-1$, respectively. In this way, all parameters in
our solution are exhausted after matching the desired flux. Therefore, apart from
the $SU(2)_R$ rotation moduli on the scalars,
we find that there are no more `moduli' in the 8 parameters that we found in
the asymptotic expansions. This implies that the ansatz we employed is insufficient
to generate the most general $U(2)\times U(2)$ monopoles, as the degeneracy from
the index seems to demand two complex moduli (to account for two factors of $SU(2)$
characters). We come back to this point later in this subsection.

\begin{figure}[t!]
  \begin{center}
    \includegraphics[width=8cm]{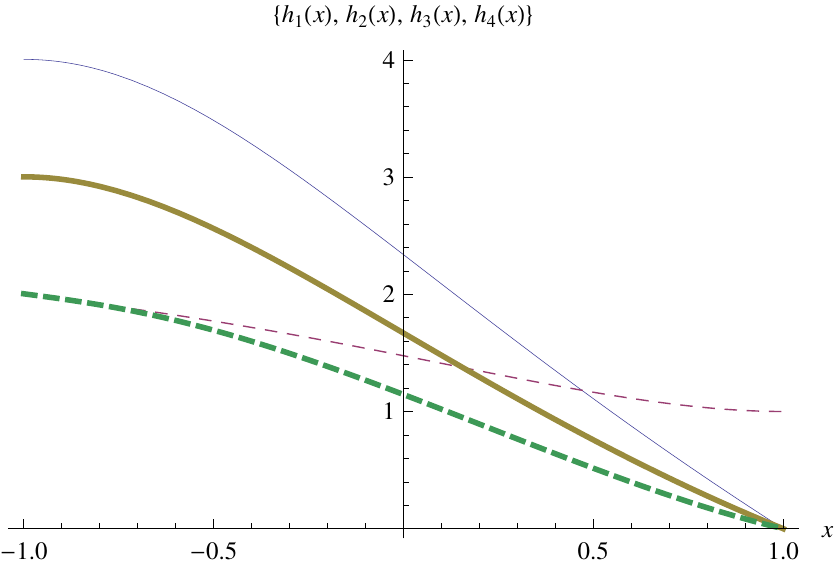}\hspace{0.5cm}
    \includegraphics[width=8cm]{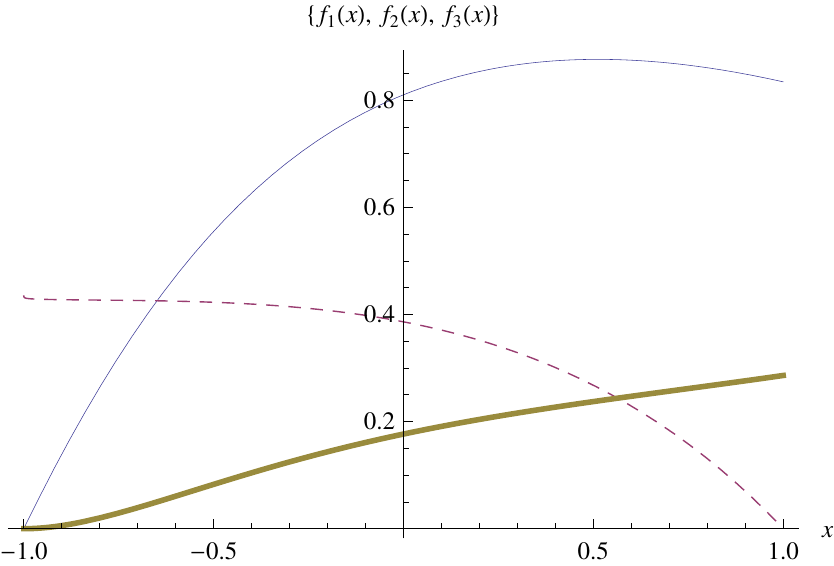}
\caption{Profiles of $h_{1,2}$, $h_{3,4}\equiv\tilde{h}_{1,2}$ and $f_{1,2,3}$ with
$H\!\cong\!(4,1)$, $\tilde{H}\!\cong\!(3,2)$. On the left side, the four curves
are for $h_1$ (thin), $h_2$ (thin dashed), $h_3$ (thick), $h_4$ (thick dashed).
On the right side, the three curves are for $f_1$ (thin), $f_2$ (thin dashed),
$f_3$ (thick), with $f_4\!=\!0$ for the ground state.}\label{general}
  \end{center}
\end{figure}
For numerical calculations, it is convenient to use the $\varphi$ dependent
gauge transformation in the overall $U(1)$ (which was unused yet) to set
$h_1(1)\!=\!0$. Fig.\ref{general} shows a profile of the functions $h_1(x)$, $h_2(x)$,
$h_3(x)\equiv\tilde{h}_1(x)$, $h_4(x)\equiv\tilde{h}_2(x)$ and $f_1(x)$, $f_2(x)$,
$f_3(x)$ (with $f_4(x)\!=\!0$) for $n_1\!=\!4.00432$, $n_2\!=\!1.00194$,
$\tilde{n}_1\!=\!3.00209$, $\tilde{n}_2\!=\!2.00417$, tuned close to the quantized monopole
charges $H\!=\!(4,1)$, $\tilde{H}\!=\!(3,2)$. We took $c\!=\!\tilde{n}_2\!-\!n_2\!=\!1$
for the ground states. We first used the above series-expanded
functions between $.9999\!<\!x\!<\!1$ to determine
the values of six functions at $x\!=\!.9999$ in terms of $a_2\!=\!g_1(1)$,
$a_3\!=\!\tilde{g}_1(1)$, $a_4$ defined above, and then obtained a numerical
solutions for $-.9999\leq x\leq.9999$. Two of these three numbers can be regarded as
determining the two fluxes apart from $\tilde{n}_2\!-\!n_2$, while one should be carefully
chosen to match to the regular asymptotic solution near $x\!=\!-1$. (One is killed by
overall $U(1)$.) This is quite a tedious trial-and-error exercise, but is doable.
For generic choice of parameters, we find that it is $g,\tilde{g}$ functions
which diverge at $x\!=\!-1$, but not $h,\tilde{h}$ functions. Thus, by observing the
changes of finite values of $h,\tilde{h}$ at $x\!=\!-1$ as we change the parameters
$a_2,a_3,a_4$, we could navigate through the parameter space to tune their values with
correctly quantized fluxes. The profiles in Fig.\ref{general} are obtained by
taking $a_2\!=\!-1.12$, $a_3\!=\!-.334$ and $a_4\!=\!-.375$.

Like the solutions with $n_2\!=\!0$ in the previous subsection, here
we can also generate more solutions by acting an $SU(2)_R$ rotation on
two scalars $\phi_1,\phi_2$ in our ansatz. There again appears two complex numbers
$b_1,b_2$ (subject to $|b_1|^2\!+\!|b_2|^2\!=\!1$) as moduli. The symplectic 2-form
now becomes
\begin{equation}
  \omega=-\frac{ik}{\pi}\left(\delta b_a\wedge\delta b_a^\ast\right)\int_{S^2}
  \left[(g_1\!-\!\tilde{g}_1)f_1+(g_2\!-\!\tilde{g}_2)f_2-(g_1\!-\!\tilde{g}_2)f_3
  -(g_2\!-\!\tilde{g}_1)f_4\right]\ .
\end{equation}
Here, inserting $f_4\!=\!0$ for the ground state solutions and using (\ref{gauss}),
one obtains
\begin{equation}
  \omega=ik(2\tilde{n}_1\!-\!n_1\!+\!n_2)\delta b_a\wedge\delta b_a^\ast\ .
\end{equation}
Again denoting by $N_a$ the occupation numbers for the two harmonic oscillators,
the ground state degeneracy from our solution is given by counting all possible
occupation numbers subject to the constraint
\begin{equation}
  N_1+N_2=k(2\tilde{n}_1\!-\!n_1\!+\!n_2)\ .
\end{equation}
Introducing the chemical potential $r$ for the Cartan charge $\pm\frac{1}{2}$
of $SU(2)_R$, partition function for our ground states is
$\chi_{k(2\tilde{n}_1\!-\!n_1\!+\!n_2)}(r)$, which is smaller than the degeneracy
(\ref{net-character}) from the index for $n_2\!\neq\!0$:
\begin{equation}
  \chi_{kn_2}(r)\chi_{k(2\tilde{n}_1\!-\!n_1)}(r)=\chi_{k(2\tilde{n}_1\!-\!n_1\!+\!n_2)}+
  \chi_{k(2\tilde{n}_1\!-\!n_1\!+\!n_2)\!-\!2}+\cdots+\chi_{k(2\tilde{n}_1\!-\!n_1\!-\!n_2)}
  >\chi_{k(2\tilde{n}_1\!-\!n_1\!+\!n_2)}\ .
\end{equation}
Here we have decomposed the product representation of $SU(2)$ into irreducible
representations, which shows that our moduli only captures the states with highest
Casimir charge. For this reason, we suspect that our ansatz is not the most general
one unless $n_2\!=\!0$. More comments are in order in the conclusion.

\begin{figure}[t!]
  \begin{center}
    \includegraphics[width=8cm]{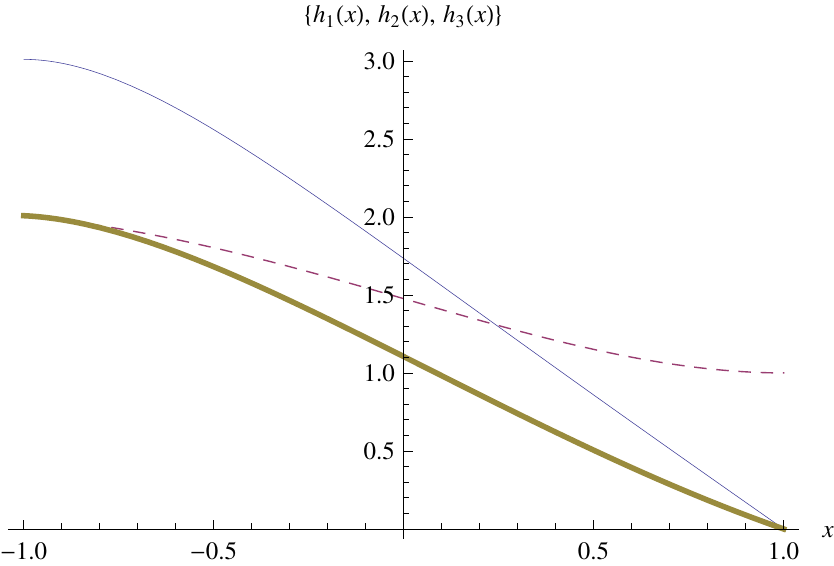}\hspace{0.5cm}
    \includegraphics[width=8cm]{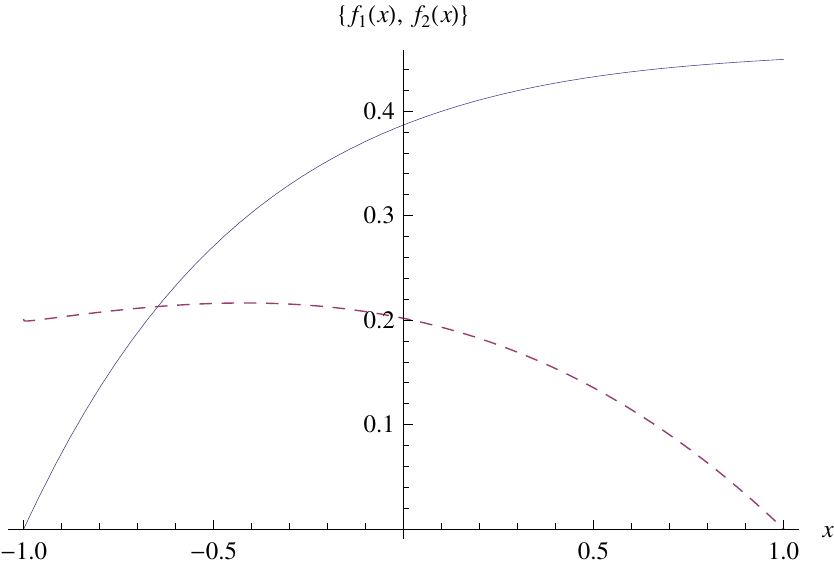}
\caption{A solution with $H\!\cong\!(3,1)$, $\tilde{H}\!\cong\!(2,2)$.
$h_1$ (thin), $h_2$ (dashed), $h_3\!\equiv\!\tilde{h}_1\!=\!\tilde{h}_2$ (thick)
on the left. $f_1\!=\!f_3$ (thin), $f_2\!=\!f_4$ (dashed) on the right.
}\label{enhanced-1}
  \end{center}
\end{figure}
\begin{figure}[t!]
  \begin{center}
    \includegraphics[width=8cm]{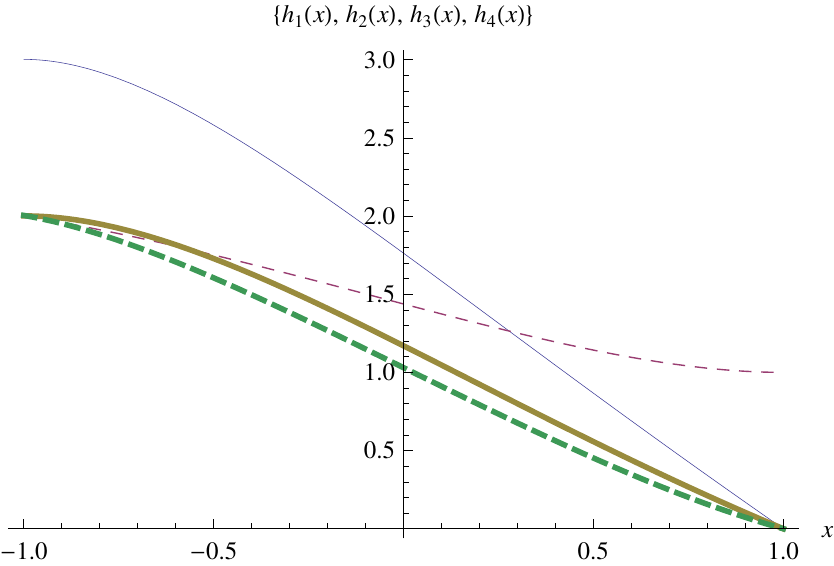}\hspace{0.5cm}
    \includegraphics[width=8cm]{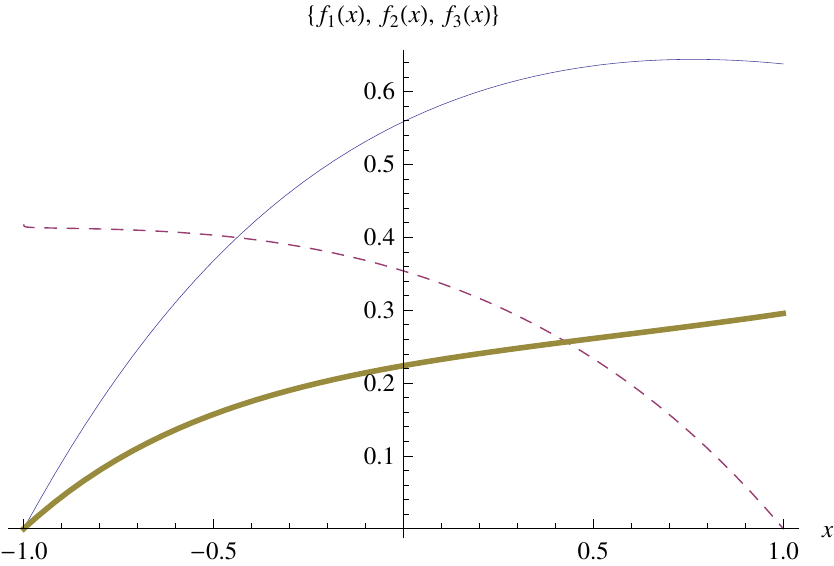}
\caption{Another solution with $H\!\cong\!(3,1)$, $\tilde{H}\!\cong\!(2,2)$.
$h_1$ (thin), $h_2$ (thin dashed), $h_3\!\equiv\!\tilde{h}_1$ (thick),
$h_4\!\equiv\!\tilde{h}_2$ (thick dashed) on the left. $f_1$ (thin),
$f_2$ (dashed), $f_3$ (thick) with $f_4\!=\!0$ on the right.}\label{enhanced-2}
  \end{center}
\end{figure}
If $\tilde{n}_1\!=\!\tilde{n}_2$, there are more solutions than the generic case with
$\tilde{n}_1\!\neq\!\tilde{n}_2$ discussed so far. This is because we no
longer have to impose $f_4=0$ for the ground states. Therefore, we have to search for
solutions with general equations (\ref{susy}), (\ref{gauss}). We expect that this
sector should exhibit more moduli than those generated by the $SU(2)_R$ action.
Rather than systematically studying this case, we simply present two different solutions
with same flux, to illustrate the presence of an extra moduli. Firstly,
we find a solution with $H\!=\!(3.00928,1.00578)$, $\tilde{H}\!=\!(2.00753,2.00753)$
satisfying $\tilde{h}_1\!=\!\tilde{h}_2$, $\tilde{g}_1\!=\!\tilde{g}_2$, $f_1\!=\!f_3$,
$f_2\!=\!f_4$. See Fig.\ref{enhanced-1}. We present another solution with same flux,
using the expansion with $f_4\!=\!0$. The solution in Fig.\ref{enhanced-2}
have fluxes $H\!=\!(3.00205,1.00245)$, $\tilde{H}\!=\!(2.00067,2.00384)$.
The presence of extra moduli is desirable since
we know from (\ref{U(2)-index}) that more states than (\ref{net-character})
are expected for $\tilde{n}_1\!=\!\tilde{n}_2$.


One can also find many `excited solutions,' where all or part of
the conditions (\ref{parameter-ground}), (\ref{zero-scalar}) are violated. We have
explicitly constructed various excited solutions, which will not be presented
here.

\subsection{Special solutions for $U(N)\times U(N)$}

In the previous subsections, we paid attention to the monopole solutions in the
$U(2)\times U(2)$ theory, which were by themselves fairly nontrivial. In this
subsection, to illustrate that similar analysis can be done for general
$U(N)\times U(N)$ theory, we present a consistent ansatz (mimicking our
$U(1)\times U(2)$ ansatz) which provides a set of ordinary differential
equations for the monopoles.

We consider a configuration which carries nonzero
$U(1)^N\times U(1)^N\subset U(N)\times U(N)$ monopole charges. The charges
that we consider take the following form,
\begin{equation}
  H={\rm diag}(n_1,n_2,\cdots,n_{N\!-\!1},0)\ ,\ \
  \tilde{H}={\rm diag}(\tilde{n}_1,\tilde{n}_2,\cdots,\tilde{n}_N)\ ,
\end{equation}
that is, with $n_N\!=\!0$. Our ansatz for the gauge fields is again diagonal,
\begin{equation}
  (A_\mu)_{ij}=\delta_{ij}A^i_\mu\ \ ({\rm with}\ (A_\mu)_{NN}\!=\!0)\ ,\ \
  (\tilde{A}_\mu)_{ij}=\delta_{ij}\tilde{A}_\mu^i\ ,
\end{equation}
where $i,j=1,2,\cdots,N$ are the indices for either factor of $U(N)\times U(N)$
as appropriate. Let us again set $A_\theta^i\!=\!\tilde{A}^i_\theta\!=\!0$, and also
take $A^i_t$, $\tilde{A}^i_t$, $A^i_\varphi$, $\tilde{A}_\varphi$ to depend only on
$\theta$. Inspired by the solutions for the $U(1)\times U(2)$ case, the
anti-bifundamental scalars are taken to be
\begin{equation}
  (\phi_1)_{ij}=\delta_{ij}\psi_i\ e^{i(m_i\varphi\!-\!\omega_it)}\ \ ({\rm with}\
  (\phi_1)_{NN}=0)\ ,\ \ (\phi_2)_{ij}=\delta_{i,j\!+\!1}\chi_i\
  e^{i(\tilde{m}_i\varphi\!-\!\tilde\omega_it)}\ ,
\end{equation}
with $2(N\!-\!1)$ complex components.
The $\varphi,t$ dependent phases can be gauged away by introducing $\varphi,t$ dependent
gauge transformation in $U(1)^N\times U(1)^N$$\subset U(N)\times U(N)$
(at least formally, without worrying about
singularities of the vector potentials near the poles).

From the supersymmetry condition and the Gauss' law, one obtains the following conditions.
Firstly, the phases of the $2(N\!-\!1)$ complex scalar components $\psi_i$, $\chi_i$ are
all constants. Then, defining $f_i\!=\!\frac{2\pi}{k}|\psi_i|^2$,
$\tilde{f}_i\!=\!\frac{2\pi}{k}|\chi_i|^2$ and
\begin{equation}
  g_i=A^i_t-\tilde{A}_t^i\ ,\ \ h_i=A^i_\varphi-\tilde{A}^i_\varphi\ ,\ \
  \tilde{g}_i=A^i_t-\tilde{A}^{i\!+\!1}_t\ ,\ \ \tilde{h}_i=A^i_\varphi-
  \tilde{A}^{i\!+\!1}_\varphi
\end{equation}
for $i=1,2,\cdots,N\!-\!1$, the conditions for the supersymmetric configurations are
\begin{eqnarray}
  &&-x(f_i\!-\!f_{i\!+\!1})=\tilde{g}_i+\tilde{h}_i+\frac{1}{2}\ \ \ ({\rm for}\ i=1,2,\cdots,N\!-\!2)
  \ ,\ \ -xf_{N\!-\!1}=\tilde{g}_{N\!-\!1}+\tilde{h}_{N\!-\!1}+\frac{1}{2}\nonumber\\
  &&-x\tilde{f}_1=g_1+h_1+\frac{1}{2}\ ,\ \ -x(\tilde{f}_{i\!+\!1}-\tilde{f}_i)=g_{i\!+\!1}+h_{i\!+\!1}
  +\frac{1}{2}\ \ \ ({\rm for}\ i=1,2,\cdots,N\!-\!2)
\end{eqnarray}
and
\begin{eqnarray}
  (1-x^2)g_1^\prime=2\tilde{h}_1\tilde{f}_1&,&(1-x^2)g_{i\!+\!1}^\prime=
  2(\tilde{h}_{i\!+\!1}\tilde{f}_{i\!+\!1}-\tilde{h}_i\tilde{f}_i)\nonumber\\
  (1-x^2)\tilde{g}_i^\prime=2(h_if_i-h_{i\!+\!1}f_{i\!+\!1})&,&
  (1-x^2)\tilde{g}_{N\!-\!1}^\prime=2h_{N\!-\!1}f_{N\!-\!1}\nonumber\\
  h_1^\prime=2\tilde{g}_1\tilde{f}_1&,&h_{i\!+\!1}^\prime=
  2(\tilde{g}_{i\!+\!1}\tilde{f}_{i\!+\!1}-\tilde{g}_i\tilde{f}_i)\nonumber\\
  \tilde{h}_i^\prime=2(g_if_i-g_{i\!+\!1}f_{i\!+\!1})&,&
  \tilde{h}_{N\!-\!1}^\prime=2g_{N\!-\!1}f_{N\!-\!1}\ ,
\end{eqnarray}
which provide $4(N\!-\!1)$ differential equations for $g_i,h_i,\tilde{g}_i,\tilde{h}_i$.
As a simple example, we can consistently set $g\!\equiv\!g_i\!=\!\tilde{g}_i$,
$h\!\equiv\!h_i\!=\!\tilde{h}_i$ and
$f\!\equiv\!f_i\!-\!f_{i\!+\!1}=f_{N\!-\!1}\!=\!\tilde{f}_1\!=\!\tilde{f}_{i\!+\!1}\!-\!\tilde{f}_i$
for all $i\!=\!1,2,\cdots,N\!-\!2$ components. The magnetic fluxes are given by
\begin{equation}
  H={\rm diag}(Nn,Nn,\cdots,Nn,0)\ ,\ \ \tilde{H}=((N\!-\!1)n,\cdots,(N\!-\!1)n)\ ,\nonumber
\end{equation}
while the differential and algebraic conditions reduce to $g\!+h\!+\frac{1}{2}=-xf$,
$h^\prime\!=\!2gf$, $(1-x^2)g^\prime\!=\!2hf$. These equations are solved for various values of
$n$ in section 2.1, which also provides new solutions of the $U(N)\times U(N)$ theory.

\section{Conclusion and discussions}

In this paper, we studied the semi-classical solutions for the
magnetic monopole operators in the $\mathcal{N}\!=\!6$
Chern-Simons-matter theory. As local operators in CFT are in 1-to-1
correspondence with the states on $S^2\times\mathbb{R}$, and as the
monopoles' energies are proportional to the Chern-Simons level $k$
which is the inverse coupling constant, the corresponding states can
be well described by classical solitonic solutions in the weak-coupling
regime (with large $k$). We paid special attention to the classical
solutions which would account for the ground states after
quantization.

One purpose of this work was to explicitly show the existence  of
monopole operators with various monopole charges. Namely, the
existence of BPS monopole operators with general $U(1)^N\times
U(1)^N$ magnetic charges $H\!=\!(n_1,n_2,\cdots,n_N)$,
$\tilde{H}\!=\!(\tilde{n}_1,\tilde{n}_2,\cdots,\tilde{n}_N)$ in
$U(N)\times U(N)$ were predicted in \cite{Kim:2009wb}. While
monopole operators with $H\!=\!\tilde{H}$ has been studied in
various ways
\cite{Berenstein:2008dc,Berenstein:2009sa,Benna:2009xd,Kim:2009ia},
those with $H\neq\tilde{H}$ have not been directly studied in the literature.
Now with our work in this paper, we can clearly see why the latter
kind of operators were relatively harder to study more directly. Monopole
operators with $H\!=\!\tilde{H}$ in the context of M2-branes and AdS/CFT
have been studied mostly in the
context of chiral operators, which do not carry spatial spins.
Therefore, the matter fields in the classical
solutions are in s-waves, which leave the magnetic fields to be
uniform on $S^2$. On the other hand, it is known that local
operators containing monopoles with $H\neq\tilde{H}$ carry spatial
spins \cite{Kim:2009wb}. Due to the lack of spherical symmetry on $S^2$,
the matter fields back-react to the magnetic field and makes the
latter non-uniform on $S^2$.

We emphasize that monopoles with such non-uniform
magnetic fields are ubiquitous in Chern-Simons-matter theories. To
demonstrate this point, we explicitly obtained monopole solutions in
a (supersymmetric) $U(1)$ Chern-Simons-matter theory coupled to one
fundamental matters, which is one of the simplest Chern-Simons-matter
theories that we can imagine. See appendix A.

We also studied the ground state degeneracies of the monopoles with
$H\neq\tilde{H}$. When one of the $U(1)^2\times U(1)^2$ magnetic
charges is zero in the $U(2)\times U(2)$ theory, all quantum numbers
and the degeneracy of ground states predicted by the index are
successfully reproduced from our classical solutions. For general
$U(2)\times U(2)$ magnetic charges, our ansatz seems to be
insufficient to obtain the known ground state degeneracy, while all
quantum numbers are correctly obtained. At the classical level, the
dimension of the moduli space in our solution is smaller than what
we expect for the most general solution. Probably a genuine
non-Abelian ansatz for the gauge fields is necesary to understand
the most general solutions. To understand this, it should be
important to investigate general BPS zero modes around our solution by
studying small fluctuations. We leave this study to the future.

It would also be interesting to study possible relation to the
vortex solitons in the mass-deformed Chern-Simons-matter theories. As monopole
operators are vortex-creating operators, BPS vortex solutions of
\cite{Kim:2009ny,Auzzi:2009es} in the mass-deformed theory on
$\mathbb{R}^{2\!+\!1}$ and our monopoles in the conformal theory on
$S^2\times\mathbb{R}$ should be two special limits of vortex-like
solitons of the mass-deformed theory on $S^2\times\mathbb{R}$. Perhaps
this relation could provide a hint towards the zero mode structures
of our monopoles from facts known for the vortices. Note that the
general study of BPS vortices in mass-deformed theories also has been
somewhat mysterious in that the true BPS vacua have not been correctly
identified. As this mystery has been resolved recently in \cite{Kim:2010mr},
it should also be interesting to revisit the study of vortices and
their roles in the gauge/gravity duality.

It will also be interesting to systematically study the ground state
degeneracy of monopole operators with various magnetic fluxes. A reason
why we feel this problem is interesting is the following. In the superconformal
index of $U(N)$ $\mathcal{N}\!=\!4$ Yang-Mills theory, it has been shown that
the large $N$ limit of the index is much smaller than the partition function
(or free energy, scaling like $N^2$ in the high temperature phase) which we
expect from the gravity dual \cite{Kinney:2005ej}. This is presumably due to a vast
cancelation between the contrbution from bosonic and fermionic states to
the index. Now with various monopole sectors in Chern-Simons-matter theories,
there exist the notion of `many ground states' for given monopole charges.
As far as we have studied in various important sectors, these ground degeneracies
from the index appear to be nonnegative and scale with a positive power of $k$.
For instance, the degeneracy from (\ref{U(2)-index}) for the $U(2)\times U(2)$
monopoles are quadratic in $k$ as the index is a multiplication of two characters.
Therefore, summing over various monopole sectors, in particular the ground states,
there could be more
nontrivial large $N$ scaling of the index free energy than 4 dimensional
theories. Especially, it should be interesting to
see whether one can get the mysterious scaling $N^{\frac{3}{2}}$. See
also \cite{Drukker:2010nc} for a recent observation of this factor in
the Chern-Simons-matter theory.

\vskip 0.5cm

\hspace*{-0.8cm} {\bf\large Acknowledgements}

\vskip 0.2cm

\hspace*{-0.75cm} We are grateful to Ki-Myeong Lee and Jaemo Park
for helpful discussions. This work is supported by the Research
Settlement Fund for the new faculty of Seoul National University
(SK), and also by the National Research Foundation of Korea (NRF) Grants
No. 2010-0007512 (SK, HK), 2007-331-C00073, 2009-0072755, 2009-0084601 (HK).

\appendix

\section{$U(1)$ monopoles with a fundamental matter}

After the discovery of $\mathcal{N}\!=\!6$ Chern-Simons-matter theories,
many generalizations with lower supersymmetry for M2-branes with AdS$_4$ duals have been
constructed. Restricting to the case with $U(N)\times U(N)$
gauge group, monopole operators with $H\!=\!\tilde{H}$ play very important roles
in various aspects. Also, these monopoles are technically more feasible to study in that
they host chiral operators with uniform magnetic fluxes. However, one of the main messages
of this paper is to emphasize that monopoles with $H\neq\tilde{H}$ are rather generic
in Chern-Simons-matter theories. Or speaking more generally without even considering
$U(N)_k\times U(N)_{-k}$ gauge group and level, monopoles with spherically non-uniform
magnetic fields and matters are generic than exceptional.

As a simple illustration, we consider an $\mathcal{N}\!=\!2,3$ supersymmetric
Chern-Simons-matter theory with $U(1)_k$ gauge group and level, coupled to one fundamental
hypermultiplet. For instance, the D-brane construction of such field theories are discussed
in \cite{Giveon:2008zn}. We simply chose this model to emphasize our point without
seriously extending any analysis of this paper, but similar solutions can be found
in different models. The bosonic Lagrangian containing the $U(1)$
gauge field $A_\mu$ and a complex anti-fundamental scalar $\phi$ (with charge $-1$) in
the hypermultiplet is given by
\begin{equation}
  \mathcal{L}=\frac{k}{4\pi}AdA-D_\mu\phi D^\mu\phi^\ast-\phi\sigma^2\phi^\ast
  -\phi D\phi^\ast,
\end{equation}
where $\sigma=-\frac{2\pi}{k}|\phi|^2$ in this truncation. We also radially quantize the
theory, giving a conformal mass to the scalar $\phi$. Taking the familiar ansatz
\begin{equation}
  A_t=g(\theta)\ ,\ \ A_\varphi=h(\theta)\ ,\ \ A_\theta=0\ ,\ \ \phi=\psi(\theta)\ ,
\end{equation}
The Gauss' law and supersymmetry condition under $Q_-$ (in the $\mathcal{N}\!=\!2$
supercharges $Q_\alpha$) are reduced to
\begin{equation}
  xf=-\left(g+h+\frac{1}{2}\right)\ ,\ \ h^\prime=2gf\ ,\ \ (1-x^2)g^\prime=2hf\ ,
\end{equation}
where $x\equiv\cos\theta$, $f\!\equiv\!\frac{2\pi}{k}|\psi|^2$ and primes again denote
$x$ derivatives. The analysis of the solutions is exactly the same as that in section 3.1.

\end{document}